\documentclass[aip,prb,unsortedaddress,superscriptaddress,reprint]{revtex4-1}
\usepackage{hyperref}
\usepackage{graphicx}
\usepackage{amsmath}
\usepackage{amssymb}
\usepackage{times,mathptmx}
\usepackage{ifthen}
\usepackage{color}
\usepackage{proof}

\newif\ifcom
\newif\ifdel
\comtrue               %
\delfalse

\begin{document}

\title{Quantitative study of the spin Hall magnetoresistance in ferromagnetic insulator/normal metal hybrids}
\author{Matthias Althammer}
\email{Matthias.Althammer@wmi.badw-muenchen.de}
\affiliation{Walther-Meissner-Institut, Bayerische Akademie der Wissenschaften, Walther-Meissner-Strasse 8, 85748 Garching, Germany}
\affiliation{University of Alabama, Center for Materials for Information Technology MINT and Dept.~Chem, Tuscaloosa, AL 35487 USA }
\author{Sibylle Meyer}
\affiliation{Walther-Meissner-Institut, Bayerische Akademie der Wissenschaften, Walther-Meissner-Strasse 8, 85748 Garching, Germany}
\author{Hiroyasu Nakayama}
\affiliation{Institute for Materials Research, Tohoku University, Sendai 980-8577, Japan}
\affiliation{Laboratory for Nanoelectronics and Spintronics, Research Institute of Electrical Communication, Tohoku University, Sendai 980-8577, Japan}
\author{Michael Schreier}
\affiliation{Walther-Meissner-Institut, Bayerische Akademie der Wissenschaften, Walther-Meissner-Strasse 8, 85748 Garching, Germany}
\author{Stephan Altmannshofer}
\affiliation{Walther-Meissner-Institut, Bayerische Akademie der Wissenschaften, Walther-Meissner-Strasse 8, 85748 Garching, Germany}
\author{Mathias Weiler}
\affiliation{Walther-Meissner-Institut, Bayerische Akademie der Wissenschaften, Walther-Meissner-Strasse 8, 85748 Garching, Germany}
\author{Hans Huebl}
\affiliation{Walther-Meissner-Institut, Bayerische Akademie der Wissenschaften, Walther-Meissner-Strasse 8, 85748 Garching, Germany}
\author{Stephan Gepr\"{a}gs}
\affiliation{Walther-Meissner-Institut, Bayerische Akademie der Wissenschaften, Walther-Meissner-Strasse 8, 85748 Garching, Germany}
\author{Matthias Opel}
\affiliation{Walther-Meissner-Institut, Bayerische Akademie der Wissenschaften, Walther-Meissner-Strasse 8, 85748 Garching, Germany}
\author{Rudolf Gross}
\affiliation{Walther-Meissner-Institut, Bayerische Akademie der Wissenschaften, Walther-Meissner-Strasse 8, 85748 Garching, Germany}
\affiliation{Physik-Department, Technische Universit\"{a}t M\"{u}nchen, 85748 Garching, Germany}
\author{Daniel Meier}
\affiliation{Fakult\"{a}t f\"{u}r Physik, Universit\"{a}t Bielefeld, 33615 Bielefeld, Germany}
\author{Christoph Klewe}
\affiliation{Fakult\"{a}t f\"{u}r Physik, Universit\"{a}t Bielefeld, 33615 Bielefeld, Germany}
\author{Timo Kuschel}
\affiliation{Fakult\"{a}t f\"{u}r Physik, Universit\"{a}t Bielefeld, 33615 Bielefeld, Germany}
\author{Jan-Michael Schmalhorst}
\affiliation{Fakult\"{a}t f\"{u}r Physik, Universit\"{a}t Bielefeld, 33615 Bielefeld, Germany}
\author{G\"{u}nter Reiss}
\affiliation{Fakult\"{a}t f\"{u}r Physik, Universit\"{a}t Bielefeld, 33615 Bielefeld, Germany}
\author{Liming Shen}
\affiliation{University of Alabama, Center for Materials for Information Technology MINT and Dept.~Chem, Tuscaloosa, AL 35487 USA }
\author{Arunava Gupta}
\affiliation{University of Alabama, Center for Materials for Information Technology MINT and Dept.~Chem, Tuscaloosa, AL 35487 USA }
\author{Yan-Ting Chen}
\affiliation{Kavli Institute of NanoScience, Delft University of Technology, 2628 CJ Delft, The Netherlands}
\author{Gerrit E. W. Bauer}
\affiliation{Institute for Materials Research, Tohoku University, Sendai 980-8577, Japan}
\affiliation{WPI Advanced Institute for Materials Research, Tohoku University, Sendai 980-8577, Japan}
\affiliation{Kavli Institute of NanoScience, Delft University of Technology, 2628 CJ Delft, The Netherlands}
\author{Eiji Saitoh}
\affiliation{Institute for Materials Research, Tohoku University, Sendai 980-8577, Japan}
\affiliation{CREST, Japan Science and Technology Agency, Tokyo 102-0076, Japan}
\affiliation{The Advanced Science Research Center, Japan Atomic Energy Agency, Tokai 319-1195, Japan}
\affiliation{WPI Advanced Institute for Materials Research, Tohoku University, Sendai 980-8577, Japan}
\author{Sebastian T. B. Goennenwein}
\email{Sebastian.Goennenwein@wmi.badw-muenchen.de}
\affiliation{Walther-Meissner-Institut, Bayerische Akademie der Wissenschaften, Walther-Meissner-Strasse 8, 85748 Garching, Germany}
\date{\today}
\begin{abstract}
We experimentally investigate and quantitatively analyze the spin Hall magnetoresistance effect in ferromagnetic insulator/platinum and ferromagnetic insulator/nonferromagnetic metal/platinum hybrid structures. For the ferromagnetic insulator we use either yttrium iron garnet, nickel ferrite or magnetite and for the nonferromagnet copper or gold. The spin Hall magnetoresistance effect is theoretically ascribed to the combined action of spin Hall and inverse spin Hall effect in the platinum metal top layer. It therefore should characteristically depend upon the orientation of the magnetization in the adjacent ferromagnet, and prevail even if an additional, nonferromagnetic metal layer is inserted between Pt and the ferromagnet. Our experimental data corroborate these theoretical conjectures.
Using the spin Hall magnetoresistance theory to analyze our data, we extract the spin Hall angle and the spin diffusion length in platinum. For a spin mixing conductance of $4\times10^{14}\;\mathrm{\Omega^{-1}m^{-2}}$ we obtain a spin Hall angle of $0.11\pm0.08$ and a spin diffusion length of $(1.5\pm0.5)\;\mathrm{nm}$ for Pt in our thin film samples.
\end{abstract}
\maketitle
\section{Introduction}
Pure spin currents - which transport only  (spin) angular momentum, but no electrical charge - represent a new paradigm for spin transport and spin electronics.
In the last few years, two spin current generation methods have evolved: the spin Seebeck effect (SSE)~\cite{uchida_electric_2011,adachi_linear-response_2011,adachi_gigantic_2010,uchida_phenomenological_2009,xiao_theory_2010,uchida_longitudinal_2010,uchida_observation_2010,uchida_observation_2008,jaworski_observation_2010,uchida_spin_2010,jaworski_spin-seebeck_2011,le_breton_thermal_2011,slachter_thermally_2010,weiler_local_2012} and the spin pumping effect~\cite{qiu_all-oxide_2012,saitoh_conversion_2006,ando_direct_2010,ando_electrically_2011,ando_photoinduced_2010,ando_inverse_2011,hoffmann_pure_2007,ando_observation_2012,czeschka_scaling_2011,uchida_surface-acoustic-wave-driven_2011,weiler_spin_2011,brataas_spin_2002}.  Both methods involve ferromagnet (FM)/nonferromagnet (NM) hybrid structures to generate and to detect the spin currents. While originally the focus was on electrically conducting FM/NM heterostructures, ferromagnetic insulators (FMI) are increasingly exploited in FMI/NM hybrids.~\cite{kajiwara_transmission_2010,uchida_spin_2010} FMIs are electrically insulating materials which exhibit long-range magnetic order, such that magnetic excitations (spin currents) can propagate in FMIs, while charge currents cannot. In this sense, FMIs allow to cleanly separate spin current from charge current effects. One of the prototype examples for a FMI compound is yttrium iron garnet (Y$_{3}$Fe$_{5}$O$_{12}$, YIG).~\cite{kajiwara_transmission_2010,uchida_spin_2010,heinrich_spin_2011}

The interplay between spin and charge transport in FM/NM devices gives rise to interesting physical phenomena. A prominent example is the spin Hall magnetoresistance (SMR) discovered recently in FMI/NM hybrids.~\cite{Nakayama2012,Chen_SMR_2013,[{Shortly before submission of the manuscript, we became aware of this publication on the arXiv: }]vlietstra_spin-hall_2013} The SMR is related to the absorption/reflection of a spin current density $\mathbf{J}_\mathrm{s}$ flowing along the direction normal to the FMI/NM interface. The spin current is generated by a charge current density $\mathbf{J}_\mathrm{q}$ in the NM layer via the spin Hall effect (SHE):~\cite{dyakonov_current-induced_1971,hirsch_spin_1999}
\begin{equation}
\mathbf{J}_\mathrm{s}=\alpha_\mathrm{SH} \left(-\frac{\hbar}{2e}\right)\mathbf{J}_\mathrm{q}\times\mathbf{s}.
\label{equ:YIG_SMR_SpinHalleffect}
\end{equation}
The total $\mathbf{J}_\mathrm{s}$ is here the direction of the spin current, while its spin polarization $\mathbf{s}$ is oriented perpendicular to $\mathbf{J}_\mathrm{s}$ and $\mathbf{J}_\mathrm{q}$. $\alpha_\mathrm{SH}=\sigma_\mathrm{SH}/\sigma$ is the spin Hall angle defined by the ratio of the spin Hall conductivity $\sigma_\mathrm{SH}$ and the electric conductivity $\sigma$ [\onlinecite{takahashi_spin_2008}] and $e$ is the positive elementary charge. In FMI/NM hybrids, the amount of spin current absorption/reflection by the FMI at the FMI/NM interface depends on the orientation of the FMI magnetization $\mathbf{M}$ with respect to the polarization $\mathbf{s}$ of the spin current. The amount of spin current reflected at the interface in turn induces a charge current via the inverse spin Hall effect (ISHE) in the NM layer. The total $\mathbf{M}$-orientation dependent spin current flow across the FMI/NM interface represents a dissipation channel for charge transport, and thus affects the resistance of the NM. The resulting spin Hall magnetoresistance of the NM is clearly discernible from a conventional anisotropic magnetoresistance~\cite{mcguire_anisotropic_1975} (AMR) type of effect, since the SMR depends on the angle between the magnetization $\mathbf{M}$ and the spin polarization $\mathbf{s}\perp\mathbf{J}_\mathrm{q}$, while the AMR depends on the angle between $\mathbf{M}$ and the charge current direction $\mathbf{J}_\mathrm{q}$.

Here, we give a detailed, quantitative analysis of the SMR in YIG/Pt, YIG/Au/Pt, and YIG/Cu/Pt samples. We furthermore report the observation of SMR in nickel ferrite (NiFe$_2$O$_4$)/Pt, and magnetite (Fe$_3$O$_4$)/Pt samples. The latter two materials are also important for SSE measurements since recently the SSE was observed in semiconducting NiFe$_2$O$_4$~\cite{Meier_2013} and weakly conducting Fe$_3$O$_4$~\cite{ramos_observation_2012}.

The paper is organized as follows: First, we explain the SMR effect in a simple phenomenological picture, followed by a quantitative model. Next, we present structural and magnetic data of the YIG/Pt hybrids, in connection with extensive magnetoresistance measurement data as a function of the magnetization orientation $\mathbf{m}$. We use our theoretical model to quantitatively analyze the experimental data and thus obtain estimates for the unknown parameters, i.e.,~the spin Hall angle $\alpha_\mathrm{SH}$ and spin diffusion length $\lambda_\mathrm{Pt}$ in Pt. Moreover, we verify that the SMR does not originate from a static equilibrium magnetic proximity effect in the non-magnetic Pt layer, since the SMR persists even when a second, diamagnetic NM is inserted between YIG and Pt. Finally, we show that the SMR effect is also observed in FMI/Pt hybrid structures using Fe$_3$O$_4$ or NiFe$_2$O$_4$ as the FMI.

\section{Theory of the spin Hall magnetoresistance}
\label{Sect_Theory_SMR}
The SMR effect stems from a combination of the spin Hall effect (SHE), which converts a charge current $\mathbf{J}_\mathrm{q}$ into a spin current $\mathbf{J}_\mathrm{s}$, and the inverse spin Hall (ISHE) effect, where a spin current $\mathbf{J}_\mathrm{s}$ induces a charge current $\mathbf{J}_\mathrm{q}$.~\cite{hirsch_spin_1999,takahashi_spin_2008}

We start with a phenomenological explanation of the SMR effect by discussing the influence of two different boundary conditions on the steady state in case of both the ordinary Hall effect (OHE) and the SHE.~\cite{hall_new_1879}
\begin{figure}[b,t]
  \includegraphics[width=85mm]{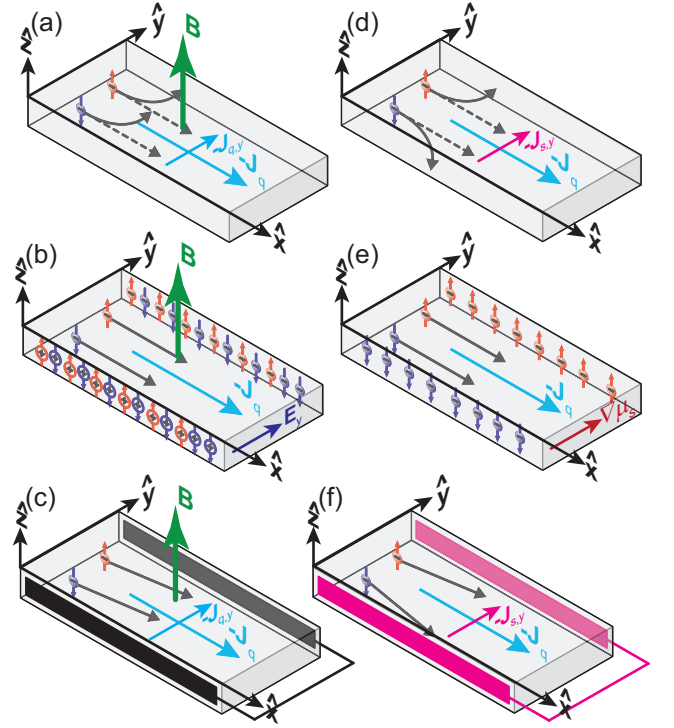}\\
  \caption[Illustration of ordinary Hall effect and spin Hall effect]{(Color online) (a) Illustration of the ordinary Hall effect in the single band model. The applied external magnetic field $\mathbf{B}\parallel\hat{\mathbf{z}}$ deflects the longitudinal charge current $\mathbf{J}_\mathrm{q}$ to one side of the sample due to the Lorentz force. (b) An open circuit boundary condition for charge transport on the transverse sides leads to a charge accumulation on the transverse sides of the sample, generating a transverse electric field $\mathbf{E}_\mathrm{y}$. It compensates the Lorentz force and leads to a steady state in which the longitudinal resistance does not depend on $\mathbf{B}$. (c) Short-circuiting the transverse sides results in a transverse current flow $\mathbf{J}_\mathrm{q,y}$. In steady state, the longitudinal resistance then depends on the applied external magnetic field. (d) Due to the spin Hall effect a charge current $\mathbf{J}_\mathrm{q}$ induces a spin current $\mathbf{J}_\mathrm{s}\parallel-\hat{\mathbf{y}}$ with a spin polarization $\mathbf{s}\parallel\hat{\mathbf{z}}$ (spin up and spin down charge carriers are deflected in opposite directions). (e) For open circuit conditions for spin transport on the transverse sides (no transverse flow of spin current) a gradient in the spin dependent electrochemical potential $\mu_\mathrm{s}$ (spin accumulation) is generated, such that the spin diffusion current compensates the spin Hall currents. In this steady state the longitudinal charge current resistance does not depend on the spin Hall effect. (f) A transverse spin current short circuit (a transverse spin potential short), in contrast, results in an additional spin current in the transverse direction. This effectively leads to a change in the longitudinal electrical resistance, due to the inverse spin Hall effect. The change in longitudinal electrical resistance between (e) and (f) is the SMR effect.}
  \label{figure:YIG_SMR_Hall_vs_SHE}
\end{figure}
We first consider the OHE in the single band model with a longitudinal charge current density $\mathbf{J}_\mathrm{q}$ in $-\hat{\mathbf{x}}$ direction as illustrated in Fig.~\ref{figure:YIG_SMR_Hall_vs_SHE}(a). Due to an external magnetic field $\mathbf{B}$ applied along $\hat{\mathbf{z}}$, the charge carriers experience a Lorentz force in the transverse ($\hat{\mathbf{y}}$) direction. In typical Hall effect measurements, the potential drop (voltage) along $\hat{\mathbf{y}}$ is recorded at zero current flow along $\hat{\mathbf{y}}$. This corresponds to the boundary condition that the charge carriers can not escape and therefore accumulate on the transverse sides (open circuit condition for charge transport). The resulting charge accumulation leads to a compensating electrical field $\mathbf{E}_\mathrm{Hall}\parallel\hat{\mathbf{y}}$, which can be detected as a voltage drop over the transverse sides. In the single band model, one finds
that the longitudinal resistance is independent of $\mathbf{B}$ (cf. Fig.~\ref{figure:YIG_SMR_Hall_vs_SHE}(b)).~\cite{ashcroft_solid_1976} In contrast, one may assume the boundary condition that $\mathbf{E}_\mathrm{Hall}=0$ (placing an electrical short between the transverse contacts as illustrated in Fig.~\ref{figure:YIG_SMR_Hall_vs_SHE}(c)), such that a transverse current $\mathbf{J}_\mathrm{q,y}$ arises, depending on the applied magnetic field magnitude $B$. This effectively reduces $\mathbf{J}_\mathrm{q}||-\hat{\mathbf{x}}$. The external current source
has to compensate for this effect by increasing the effective longitudinal voltage $V_\mathrm{long}$ applied to the sample, such that the charge current $I_\mathrm{q}$ and thus $\mathbf{J}_\mathrm{q} \parallel-\hat{\mathbf{x}}$ stays constant. For this boundary condition, the observed longitudinal resistance $R_\mathrm{long}=V_\mathrm{long}/I_\mathrm{q}$ thus does depend on the external magnetic field.~\cite{ashcroft_solid_1976}

We now turn to the SHE illustrated in Fig.~\ref{figure:YIG_SMR_Hall_vs_SHE}(d), by which the longitudinal charge current density $\mathbf{J}_\mathrm{q}$ induces a transverse spin current $\mathbf{J}_\mathrm{s}$. Since we are here concerned with the spin current flow across the FMI/NM interface, we can restrict the discussion to $-\mathbf{J}_\mathrm{s}$ along the interface normal $\hat{\mathbf{y}}$. According to Eq.\eqref{equ:YIG_SMR_SpinHalleffect}, this implies that the spin orientation $\mathbf{s}||\hat{\mathbf{z}}$. The vector $\mathbf{J}_\mathrm{s}$ thus represents a flow of $\hat{\mathbf{z}}$-polarized spins along $-\hat{\mathbf{y}}$. For an open (spin current) circuit boundary condition as illustrated in Fig.~\ref{figure:YIG_SMR_Hall_vs_SHE}(e), we obtain a spin accumulation on the transverse sides of the sample.~\cite{kato_observation_2004} This spin accumulation leads to a gradient in the spin dependent electrochemical potential $\mu_\mathrm{s}$, which compensates the spin current generated by the SHE. In the steady state, the longitudinal resistance is thus independent of the SHE, in analogy to the OHE in Fig.~\ref{figure:YIG_SMR_Hall_vs_SHE}(b). In contrast, if we consider short-circuited boundary conditions for the spin channel (shorting $\nabla \mu_\mathrm{s}$) a spin current flows through the lateral faces. The current source driving the charge current again must compensate for this transverse current flow, which results in an increase in the longitudinal resistance. This change in longitudinal resistance is the SMR effect. Dyakonov predicted a magnetoresistance, stemming from the magnetic field-dependent dephasing of a spin accumulation generated by the spin Hall effect at the edges of single NM films.~\cite{dyakonov_magnetoresistance_2007} The spin accumulation at the edges and the corresponding magnetoresistance is negligibly small for extended thin films. In contrast, we consider here the spin accumulations on the surface of the conducting film and modulate them by a FMI rather than an applied magnetic field.

Interestingly, FMI/NM heterostructures allow to gradually switch between open and closed spin current circuit boundary conditions by changing the magnetization orientation in the FMI with respect to $\mathbf{s}$. To realize an ideal spin current short circuit condition as depicted in Fig.~\ref{figure:YIG_SMR_Hall_vs_SHE}(f), FMI/NM/FMI multilayer structures are necessary. However, the SMR effect also occurs in ''simple'' FMI/NM hybrid structures, if the spin flip length in the NM is finite.
\begin{figure}[b,t]
  \includegraphics[width=85mm]{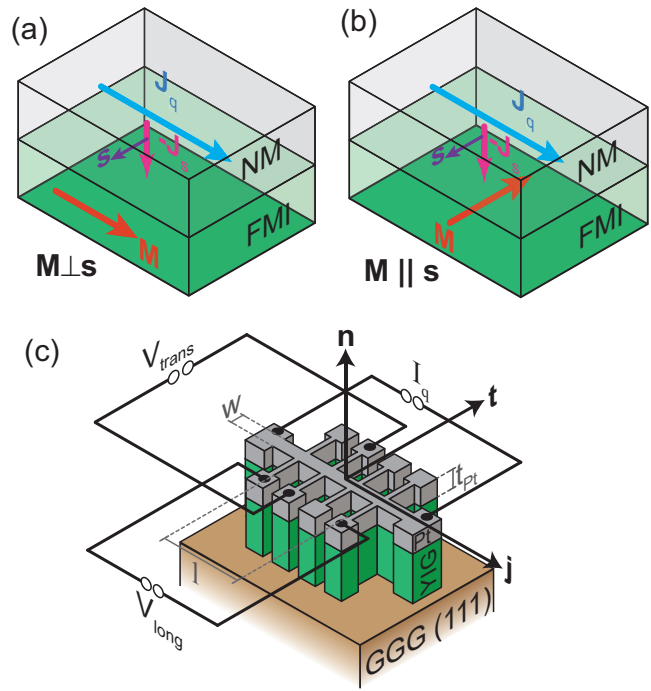}\\
  \caption[Illustration of the spin Hall magnetoresistance for YIG/Pt heterostructures]{(Color online) Graphical illustration of the SMR exhibited by FMI/NM hybrid structures. A charge current $\mathbf{J}_\mathrm{q}$ (blue arrow) flowing in the NM is converted via the SHE into a spin current $\mathbf{J}_\mathrm{s}$ (magenta arrow) flowing towards the FMI/NM interface. Due to the SHE the spin polarization $\mathbf{s}$ (violet arrow) is perpendicular to $\mathbf{J}_\mathrm{q}$ and $\mathbf{J}_\mathrm{s}$. At the FMI/NM interface the spin current is absorbed or reflected depending on the relative orientation of the magnetization $\mathbf{M}$ in the FMI (red arrow) to $\mathbf{s}$. The panels (a)-(b) show the two principle orientations of $\mathbf{M}$ with respect to the spin polarization $\mathbf{s}$: If $\mathbf{M}$ is perpendicular to $\mathbf{s}$ it is possible to transfer angular momentum via the spin torque effect, and the spin current in the NM gets absorbed by the FMI. For a collinear alignment between $\mathbf{M}$ and $\mathbf{s}$, no angular momentum transfer is possible and the spin current is reflected at the interface, leading to a spin accumulation. (c) Definition of the coordinate system defined by $\mathbf{j}$, $\mathbf{t}$, and $\mathbf{n}$ in our YIG/Pt hybrid structures. In addition, the measurement scheme used for the determination $I_\mathrm{q}$, $V_\mathrm{long}$, and $V_\mathrm{trans}$, with the Hall bar geometries $w$, $l$, and $t_\mathrm{Pt}$ is shown in the drawing.}
  \label{figure:YIG_SMR_Introduction}
\end{figure}
We consider first a charge current density $\mathbf{J}_\mathrm{q}$ flowing along $\mathbf{j}$ (Fig.~\ref{figure:YIG_SMR_Introduction}(c)) through the NM layer as indicated by the blue arrow in Fig.~\ref{figure:YIG_SMR_Introduction}(a). This charge current induces a spin current $\mathbf{J}_\mathrm{s}\parallel\mathbf{n}$ due to the SHE in the NM layer, which flows across the NM/FMI interface, depicted as the magenta arrow in Fig.~\ref{figure:YIG_SMR_Introduction}(a). The spin orientation $\mathbf{s}$ of $\mathbf{J}_\mathrm{s}$ is oriented perpendicular to $\mathbf{J}_\mathrm{s}$ and $\mathbf{J}_\mathrm{q}$ because of the SHE (cf.~Eq.(\ref{equ:YIG_SMR_SpinHalleffect})).
At the FMI/NM interface, this spin current can exert a torque on and thus be absorbed by the ferromagnet if $\mathbf{s}$ is not parallel to the magnetization $\mathbf{M}$ of the ferromagnet. This absorption can only occur in a non-collinear configuration of $\mathbf{M}$ and $\mathbf{s}$, since it only then is possible to transfer spin angular momentum from the spin current in the NM onto the magnetization of the FMI. In the Cartesian coordinate system defined by the unit vectors of the current direction ($\mathbf{j}$), the surface normal ($\mathbf{n}$), and the  transverse direction ($\mathbf{t}=\mathbf{n}\times\mathbf{j}$, see Fig.~\ref{figure:YIG_SMR_Introduction}(c)) two different configurations are possible as illustrated by Fig.~\ref{figure:YIG_SMR_Introduction}(a) and (b). In Fig.~\ref{figure:YIG_SMR_Introduction}(a) the magnetization $\mathbf{M}||\mathbf{J}_\mathrm{q}$  is oriented along $\mathbf{j}$ and thus perpendicular to the spin polarization $\mathbf{s}$ of $\mathbf{J}_\mathrm{s}$ such that the spin current is absorbed. Figure~\ref{figure:YIG_SMR_Introduction}(b) illustrates the case when $\mathbf{M}$ is oriented along $\mathbf{t}$ and thus collinear to $\mathbf{s}$. In this case, the spin current cannot be absorbed by the FMI and is reflected at the interface. In the aforementioned Cartesian coordinate system, there is a third configuration where $\mathbf{M}$ is oriented along $\mathbf{n}$. This case is analogous to Fig.~\ref{figure:YIG_SMR_Introduction}(a), since $\mathbf{M}$ and $\mathbf{s}$ are again oriented perpendicular to each other, resulting in an absorption of the spin current. In this simple picture we expect a higher resistance of the NM layer if the spin current is absorbed at the FMI/NM interface, and a lower resistance if not, such that $R_\mathrm{\mathbf{M}\perp\mathbf{s}}>R_\mathrm{\mathbf{M}\parallel\mathbf{s}}$.

For a quantitative understanding of the SMR we need to calculate the spin diffusion process in a FMI/NM bilayer system and take into account the magnetization orientation dependence of the boundary conditions at the FMI/NM interface.~\cite{Chen_SMR_2013} These calculations are based on our first simplified theory presented in the supplements of our previous publication,~\cite{Nakayama_Suppl2012} and show that the resulting steady state charge current density $\mathbf{J}_\mathrm{q,final}$ is composed of two contributions: one along the initial current direction $\mathbf{j}$ and a second part along the transverse direction $\mathbf{t}$. This leads to a magnetoresistance effect in the NM, which is sensitive to the magnetization direction of the FMI. The effect can be parameterized as a longitudinal $\rho_\mathrm{long}=V_\mathrm{long}\: w \:t_\mathrm{Pt}\:(l\: I_\mathrm{q})^{-1}$ and transverse $\rho_\mathrm{trans}=V_\mathrm{trans}\: t_\mathrm{Pt}\:(I_\mathrm{q})^{-1}$ resistivity (cf.~Fig.~\ref{figure:YIG_SMR_Introduction}(c)) of the NM as a function of magnetization orientation $\mathbf{m}=\mathbf{M}/M_\mathrm{sat}$ of the FMI~\cite{limmer_angle-dependent_2006,limmer_advanced_2008}
\begin{align}
\rho_\mathrm{long}&=\rho_0+\rho_1 m_\mathrm{t}^2,
\label{equ:YIG_SMR_rho_long}\\
\rho_\mathrm{trans}&=\rho_2 m_\mathrm{n}+\rho_3 m_\mathrm{j}m_\mathrm{t}.
\label{equ:YIG_SMR_rho_trans}
\end{align}
According to the theoretical SMR model, $\rho_1=-\rho_3$. $\left| \rho_1/\rho_0 \right|$ is the SMR effect. $\rho_2$ is a Hall-effect type resistivity. In addition to the conventional (ordinary) Hall effect of the Pt layer, $\rho_2$ also contains a Hall-type SMR contribution.~\cite{Chen_SMR_2013} While $\rho_0$, $\rho_1$, and $\rho_3$ are expected to be independent of the external magnetic field magnitude, the Pt OHE coefficient contribution to $\rho_2$ linearly changes~\cite{hall_new_1879} with $\mathbf{B}$, such that $\rho_2=\rho_{2,\mathrm{SMR}}+\rho_{2,\mathrm{OHE}}(B)$. The magnetization parameters $m_\mathrm{j}$, $m_\mathrm{t}$ and $m_\mathrm{n}$ are projections of the magnetization orientation unit vector $\mathbf{m}$ onto the coordinate system given by the unit vectors $\mathbf{j}$, $\mathbf{t}$, and $\mathbf{n}$ as illustrated in Fig.~\ref{figure:YIG_SMR_Introduction}(c).

Following the calculations detailed in Ref.~\onlinecite{Chen_SMR_2013}, the relative magnitude of the SMR effect is given by the ratios
\begin{widetext}
\begin{align}
\frac{-\rho_1}{\rho_0}&=\frac{\alpha_\mathrm{SH}^2\left(2\lambda_\mathrm{NM}^2\rho_\mathrm{NM}\right)(t_\mathrm{NM})^{-1} G_r
\tanh^2\left(\frac{t_\mathrm{NM}}{2\lambda_\mathrm{NM}}\right)}
{1+2\lambda_\mathrm{NM}\rho_\mathrm{NM}G_r\coth\left(\frac{t_\mathrm{NM}}{\lambda_\mathrm{NM}}\right)},
\label{equ:SMR_Quan_Ratio}\\
\frac{-\rho_{2,\mathrm{SMR}}}{\rho_0}&=\frac{\alpha_\mathrm{SH}^2\left(2\lambda_\mathrm{NM}^2\rho_\mathrm{NM} \right) (t_\mathrm{NM})^{-1} G_i \tanh^2\left(\frac{t_\mathrm{NM}}{2\lambda_\mathrm{NM}}\right)}{\left(1+2\lambda_\mathrm{NM}\rho_\mathrm{NM} G_r \coth \left(\frac{t_\mathrm{NM}}{\lambda_\mathrm{NM}}\right)\right)^2+\left(2\lambda_\mathrm{NM}\rho_\mathrm{NM} G_i \coth \left(\frac{t_\mathrm{NM}}{\lambda_\mathrm{NM}}\right)\right)^2}.
\label{equ:AHE_SMR_Quan_Ratio}
\end{align}
\end{widetext}
Here, $\lambda_\mathrm{NM}$ is the spin diffusion length in the NM, $\rho_\mathrm{NM}$ the resistivity of the NM, $t_\mathrm{NM}$ the thickness of the NM, and $G_r$ the real part and $G_i$ the imaginary part of the spin mixing interface conductance. The relative magnitude of this effect, i.e., the SMR magnitude $\left| \rho_1/\rho_0 \right|$, is essentially determined by $\alpha_\mathrm{SH}^2$. Taking the literature value $\alpha_\mathrm{SH}=0.012$ for NM = Pt (Ref.~\onlinecite{mosendz_detection_2010}), one thus expects a $10^{-4}$ relative resistance change for Pt. Moreover, the SMR effect will be large only if the thickness $t_\mathrm{NM}$ of the NM layer does not substantially exceed the spin diffusion length $\lambda_\mathrm{NM}$ in the NM. Last but not least, the SMR magnitude also will characteristically depend on
the resistivity $\rho_\mathrm{NM}$ and the spin mixing interface conductance $G_r$, at least as long as $2\lambda_\mathrm{NM}\rho_\mathrm{NM}G_r\leq1$.

It is important to compare Eqs.(\ref{equ:YIG_SMR_rho_long}) and (\ref{equ:YIG_SMR_rho_trans}) to the anisotropic magnetoresistance (AMR) effect in conventional electrically conductive FMs~\cite{limmer_angle-dependent_2006}:
\begin{align}
\rho_\mathrm{long}&=\rho_0+\Delta\rho m_\mathrm{j}^2,
\label{equ:Poly_AMR_rho_long}\\
\rho_\mathrm{trans}&=\rho_2 m_\mathrm{n}+\Delta\rho m_\mathrm{j}m_\mathrm{t}.
\label{equ:Poly_AMR_rho_trans}
\end{align}
At first glance, AMR and SMR appear very similar. In particular, the SMR is not discernible from the AMR of a polycrystalline ferromagnetic conductor if the  magnetization resides in the ferromagnetic film plane (i.e., the $\mathbf{j}$-$\mathbf{t}$ plane). This directly follows from the unity relation $1=m_\mathrm{j}^2+m_\mathrm{t}^2+m_\mathrm{n}^2$, as we can rewrite $m_\mathrm{j}^2$ into $1-m_\mathrm{t}^2$, for a magnetization oriented in the film plane ($m_\mathrm{n}=0$). In contrast, if $\mathbf{M}$ has an out-of-plane component ($m_\mathrm{n}\neq0$), we expect a different behavior: the magnetization orientation dependence of the SMR vanishes for a rotation of the magnetization in the plane enclosed by $\mathbf{j}$ and $\mathbf{n}$ (oopt geometry, cf.~Fig.~\ref{figure:YIG_SMR_ADMR}(c)) since then $m_\mathrm{t}=0$. In contrast, the SMR depends on $\mathbf{m}$ for a rotation of $\mathbf{m}$ in the plane enclosed by $\mathbf{t}$ and $\mathbf{n}$ (oopj geometry in Fig.~\ref{figure:YIG_SMR_ADMR}(b)) ($m_\mathrm{j}=0$). For the conventional AMR of a polycrystalline FM the situation is reversed: no $\mathbf{m}$ dependence of the AMR in the $\mathbf{t}$-$\mathbf{n}$ rotation plane (oopj geometry) and a clear $\mathbf{m}$ dependence in the $\mathbf{j}$-$\mathbf{n}$ rotation plane (oopt geometry). The SMR thus differs qualitatively from an AMR.

In our experiments we measure the magnetoresistance as a function of the magnetization orientation $\mathbf{m}$. More precisely, in these so-called angle-dependent magnetoresistance (ADMR) experiments, one records $\rho_\mathrm{long}$ and $\rho_\mathrm{trans}$ as a function of the orientation of the externally applied magnetic field $\mathbf{h}=\mathbf{H}/\left|\mathbf{H}\right|$, while maintaining a fixed magnetic field magnitude $H_\mathrm{meas}=\mathrm{const.}$. In the very same fashion as in ADMR experiments in (Ga,Mn)As thin films,~\cite{bihler_ga_1xmn_xas/piezoelectric_2008,limmer_advanced_2008,limmer_angle-dependent_2006} we then use our SMR model equations to extract the $\rho_0$, $\rho_1$, $\rho_2$, $\rho_3$ parameters from the experimental data. This requires the knowledge of $\mathbf{m}$ for all field orientations $\mathbf{h}$ and field strengths $H_\mathrm{meas}$. To describe the magnetization orientation in the FMI we use the free enthalpy approach, with the expression
\begin{equation}
G_{M}(\mathbf{m})=-\mu_0 H (\mathbf{h}\cdot \mathbf{m})+B_\mathbf{n}m_\mathrm{n}^2,
\label{equ:YIG_SMR_Free_energy}
\end{equation}
for the free enthalpy $G_{M}(\mathbf{m})$, which takes an effective shape anisotropy contribution ($B_\mathbf{n}$) and the Zeeman energy with the external magnetic field direction into account. We intentionally neglect the (small) crystalline anisotropy of our YIG samples, since we only consider experimental data taken at magnetic fields much larger than the corresponding crystalline anisotropy fields. This allows to keep the number of free modeling parameters to a minimum. Using a numeric minimization algorithm, we determine the global minimum of the free enthalpy and assume that $\mathbf{M}$ always points along this orientation. Please note that the free enthalpy in Eq.(\ref{equ:YIG_SMR_Free_energy}) is normalized to the saturation magnetization $M_\mathrm{s}$.

\section{Fabrication of FMI/NM hybrids\label{section_fabrication}}
If not explicitly indicated otherwise, the samples used in the experiments are YIG/Pt or YIG/\-NM/\-Pt heterostructures grown on (111)-oriented gadolinium gallium garnet (GGG) or yttrium aluminium garnet (YAG) substrates (cf.~Table~\ref{table:YIG_SMR_SimResults}). The YIG films were epitaxially grown via laser-MBE from a stoichiometric polycrystalline YIG target, utilizing a KrF excimer laser with a wavelength of $248\;\mathrm{nm}$ at a repetition rate of $10\;\mathrm{Hz}$.~\cite{gross_heteroepitaxial_2000,opel_spintronic_2012}  A growth optimization yields the following optimum set of YIG growth parameters: $550^\circ\mathrm{C}$ substrate temperature during deposition, $2\;\mathrm{J/cm^2}$ energy density at the target, and an oxygen atmosphere of $25\;\mathrm{\mu bar}$. The NM layers were deposited in-situ, without breaking the vacuum, on top of the YIG film using electron beam evaporation, at room temperature.

\begin{figure}[t,b]
  \includegraphics[width=85mm]{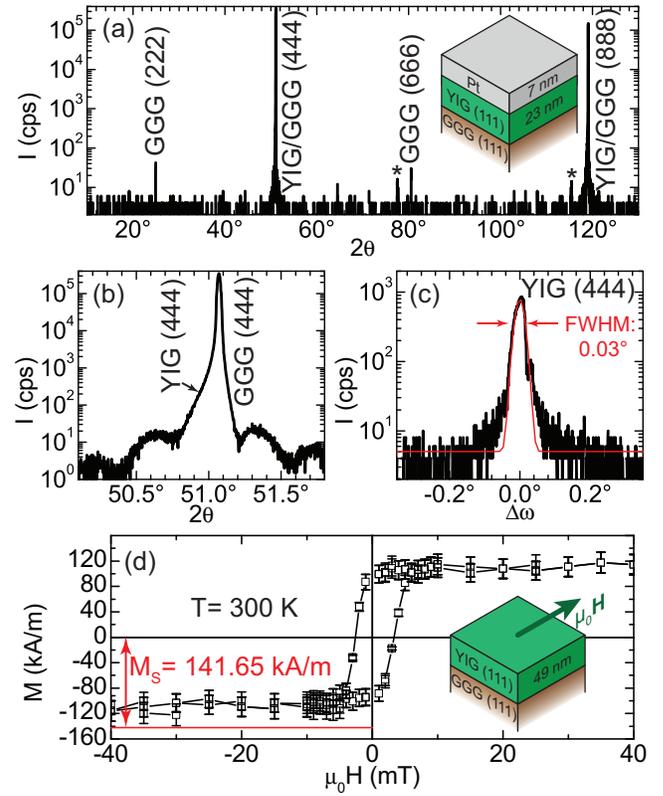}\\
  \caption{(Color online) Structural and magnetic properties of the laser-MBE grown YIG/Pt hybrids. (a) $2\theta-\omega$ scan of a YIG (23 nm)/Pt (7 nm) hybrid grown on a (111)-oriented GGG substrate. The two reflections marked with an asterisk ($\star$) are imperfections present in the substrate prior to deposition. (b) The enlargement of the $2\theta-\omega$ scan in (a) reveals satellites due to Laue oscillations around the YIG (444) reflection, indicating a coherent growth. (c) The small full width at half maximum (FWHM) of $0.03^\circ$ of the rocking curve of the YIG (444) reflection confirms the low mosaic spread of the YIG film and thus its excellent structural properties. (d) In-plane magnetization hysteresis curve of a 49 nm thick YIG layer on a GGG substrate determined from SQUID-magnetometry at $T=300\;\mathrm{K}$. The saturation magnetization $M_\mathrm{s}=110\;\mathrm{kA/m}$ of the YIG layer is lower than the bulk value of $M_\mathrm{s}=141.65\;\mathrm{kA/m}$. The large noise is due to the large paramagnetic background signal from the GGG substrate (which was subtracted from the data in panel (d)).}
  \label{LaserMBEYIG_Pt}
\end{figure}

After deposition, the structural and magnetic properties were investigated using high-resolution X-ray diffractometry (HRXRD) in a four circle diffractometer with monochromatic Cu K$\alpha_1$ radiation, as well as superconducting quantum interference device (SQUID) magnetometry. The results obtained for YIG/Pt hybrids grown on GGG substrates are compiled in Fig.~\ref{LaserMBEYIG_Pt}(a)-(d).
Figure~\ref{LaserMBEYIG_Pt}(a) shows the sharp reflections of the GGG substrate and the laser-MBE grown YIG film in a $2\theta$-$\omega$-scan. The two low intensity reflections marked by asterisks ($\star$) are due to substrate impurities. The YIG reflections, indicated by the arrow in Fig.~\ref{LaserMBEYIG_Pt}(b) for the (444) plane, are only weakly discernible from the high intensity substrate reflection due to the small lattice mismatch of only $0.03\%$ between YIG and GGG. The high structural quality of the YIG layer is not influenced by the additional deposition of a Pt film as evident from Laue oscillations visible in the vicinity of the GGG (444) reflection in Fig.~\ref{LaserMBEYIG_Pt}(b), which indicate a coherent, (111)-oriented growth of the YIG layer on GGG. From the position of the (444) reflection we calculate an interplane spacing of $d_{444}=0.1787\;\mathrm{nm}$ from the Bragg equation. Assuming an undistorted cubic lattice structure this leads to a lattice constant of $a=1.238\;\mathrm{nm}$, which is identical to the value $a_\mathrm{bulk}$ of bulk YIG.~\cite{geller_crystal_1957}
Our result $a=a_\mathrm{bulk}$ is in contrast to published data on PLD grown YIG films.~\cite{manuilov_submicron_2009,krockenberger_solid_2008,krockenberger_layer-by-layer_2009,kahl_pulsed_2003} In these publications much larger lattice constants for the YIG layer have been reported, which result from a large rhombohedral distortion of the YIG lattice due to a deficiency of iron ions in the YIG structure.~\cite{manuilov_submicron_2009,dumont_superexchange_2005} The iron deficiency can be tuned by varying the oxygen partial pressure during deposition.~\cite{dumont_superexchange_2005} From this we conclude, that our growth parameters allow the deposition of highly stoichiometric YIG films from a stoichiometric polycrystalline target.

To evaluate the mosaic spread of the laser-MBE grown YIG films we recorded X-ray rocking curves of the YIG (444) reflection for our YIG film on GGG, one typical result is displayed in Fig.~\ref{LaserMBEYIG_Pt}(c). From a Gaussian fit (red line) to the data we extract a FWHM of $0.03^\circ$. This is an excellent value for laser-MBE grown thin films and confirms the high structural quality of our samples.

In the full range of the $2\theta$-$\omega$-scan (Fig.~\ref{LaserMBEYIG_Pt}(a))
we do not observe any reflections which can be attributed to the Pt layer. We therefore assume that our Pt thin films are polycrystalline and grow without a preferential texture on YIG. We evaluated the surface roughness of the laser-MBE grown YIG/Pt hybrids on GGG substrates via X-ray reflectometry and obtain from a simulation on average $0.7\;\mathrm{nm}$ for the surface roughness amplitude of the YIG film and $0.8\;\mathrm{nm}$ for the Pt layer.

The YIG thin films deposited on diamagnetic, (111)-oriented YAG substrates (lattice mismatch: $3\%$) show similar structural properties. The $2\theta$-$\omega$-scans do not reveal any secondary phases. The FWHM of the YIG(444) rocking curves is about $0.1^\circ$ indicating a larger mosaic spread as compared to the films on lattice-matched GGG substrates. The X-ray reflectometry analysis gives on average a surface roughness amplitude of $0.8\;\mathrm{nm}$ for the YIG film and $0.9\;\mathrm{nm}$ for the Pt layer

Taken together, our laser-MBE grown YIG/Pt heterostructures exhibit excellent structural properties. Moreover, the structure of the YIG films is not influenced by the deposition of a Pt top layer.

The magnetic properties of our laser-MBE grown YIG films were analyzed using SQUID magnetometry.
A hysteresis curve at $T=300\;\mathrm{K}$ is shown in Fig.~\ref{LaserMBEYIG_Pt}(d), with the external magnetic field $H$ applied in the film plane. We subtracted the paramagnetic background signal of the GGG substrate and normalized the remaining magnetic moment to the volume of the 49 nm thick YIG layer to obtain the magnetization $M$ of the YIG film.
The $M(H)$ hysteresis curve exhibits a low coercive field $\mu_0 H_\mathrm{c}=3\;\mathrm{mT}$ and reaches a saturation magnetization $M_\mathrm{S}=110\;\mathrm{kA/m}$, which is approximately 80\% of the reported bulk value $M_\mathrm{S}=141.65\;\mathrm{kA/m}$ of YIG.~\cite{hansen_saturation_1974} We note that due to the large paramagnetic background signal the error for the determination of the saturation magnetization is large (at least 10\%) and thus alone might account for the measured difference to the bulk value (cf.~error bars in Fig.~\ref{LaserMBEYIG_Pt}(d)). Compared to the results reported by other groups,~\cite{krockenberger_solid_2008,krockenberger_layer-by-layer_2009,kahl_pulsed_2003,manuilov_submicron_2009,dorsey_epitaxial_1993,sun_growth_2012} we can confirm that the saturation magnetization of YIG thin films on GGG substrates is close to the bulk value. However, the reported coercive fields in Refs.~\onlinecite{kahl_pulsed_2003,manuilov_submicron_2009,dorsey_epitaxial_1993} at $T=300\;\mathrm{K}$ are much lower (below $1\;\mathrm{mT}$) than the one we observe in our YIG films (above $2\;\mathrm{mT}$). On the other hand, the films described in Refs.~\onlinecite{krockenberger_solid_2008,krockenberger_layer-by-layer_2009} exhibit coercive fields as large as $15\;\mathrm{mT}$. From these variations in the literature values for coercive fields and keeping in mind that $H_\mathrm{c}$ sensitively depends on the domain configuration, we conclude that the growth conditions greatly influence this quantity.

For the YIG layers on (111)-oriented, diamagnetic YAG substrates, we obtain a saturation magnetization of $M_\mathrm{S}=130\;\mathrm{kA/m}$ at $T=300\;\mathrm{K}$, close to the bulk value of YIG. The extracted coercive field $\mu_0 H_\mathrm{c}=5\;\mathrm{mT}$ at $T=300\;\mathrm{K}$ for these samples is slightly larger than the values obtained for YIG films on GGG substrates. Nevertheless, these magnetic properties confirm the state-of-the-art quality of our laser-MBE grown YIG thin films.

An additional set of YIG/Pt heterostructures has been fabricated by sputtering a Pt layer on a commercially available (FDK Corporation), $1.3\;\mathrm{\mu m}$ thick YIG layer deposited on (111)-oriented GGG substrates via liquid phase epitaxy. The YIG film was grown under a PbO-B$_2$O$_3$ flux at around $1200\;\mathrm{K}$.
For our liquid phase epitaxy prepared YIG/Pt heterostructures we find in XRD studies a FWHM of the rocking curve for the YIG (444) reflection that is as narrow as the GGG (444) substrate reflection. Magnetization measurements on our liquid phase epitaxy YIG films yield $M_\mathrm{S}=140\;\mathrm{kA/m}$ at $T=300\;\mathrm{K}$ close to the reported bulk value of YIG.~\cite{hansen_saturation_1974}

The Fe$_3$O$_4$/Pt hybrid structure used in our experiments was also fabricated by laser-MBE and electron beam evaporation. First the magnetite layer was grown via laser-MBE on a (001)-oriented MgO substrate using a polycrystalline Fe$_3$O$_4$ target, with the following set of deposition parameters:~\cite{reisinger_hall_2004,venkateshvaran_epitaxial_2009} $320^\circ\mathrm{C}$ substrate temperature during deposition, $3.1\;\mathrm{J/cm^2}$ energy density at the target, and an argon atmosphere of $0.7\;\mathrm{\mu bar}$. After the magnetite growth, the Pt layer was deposited in-situ on top of the Fe$_3$O$_4$ film using electron beam evaporation, at room temperature. Structural characterization via HRXRD showed that no secondary phases were present in the samples. Satellites due to Laue oscillations around the Fe$_3$O$_4$ (004) reflection and a narrow rocking curve of the (004) film reflection with a FWHM of $0.04^\circ$ indicate the excellent structural quality of these magnetite films.~\cite{reisinger_hall_2004,venkateshvaran_epitaxial_2009} SQUID magnetometry measurements~\cite{reisinger_hall_2004,venkateshvaran_epitaxial_2009} on these samples yield $M_\mathrm{S}=450\;\mathrm{kA/m}$ close to the bulk value of $470\;\mathrm{kA/m}$ at $T=300\;\mathrm{K}$ (Ref.~\onlinecite{magnetite_Landolt}). A Verwey transition at $T=117\;\mathrm{K}$ is visible indicating the perfect stoichiometry of the Fe$_3$O$_4$ thin film.

The nickel ferrite (NiFe$_{2}$O$_{4}$) layer was grown via direct liquid injection chemical vapor deposition on (001)-oriented MgAl$_{2}$O$_{4}$~\cite{li_growth_2011} at a substrate temperature of $700^\circ\mathrm{C}$ using Ni(acac)$_2$ and Fe(acac)$_3$ (acac=acetylacetonate) as precursors. The Pt layer was sputtered ex-situ, after cleaning the surface of the nickel ferrite by Ar ion beam milling, at room temperature. The rocking curve of the NiFe$_{2}$O$_{4}$ (004) reflection exhibited a FWHM of $0.2^\circ$ measured via HRXRD. The magnetic properties were evaluated via an alternating gradient magnetometer and yield a saturation magnetization~\cite{li_growth_2011} of $280\;\mathrm{kA/m}$ (bulk value~\cite{hellwege_6.1.2.1_1970} $M_\mathrm{S}=300\;\mathrm{kA/m}$).
\section{ADMR experiments on YIG/Pt hybrids}
\label{Sect_Experiment}
For the magnetotransport experiments we patterned Hall bar mesa structures out of the plain, laser-MBE grown YIG/Pt films using photolithography and argon ion beam milling. The width $w$ and length $l_\mathrm{HB}$ of the Hall bar were $80\;\mathrm{\mu m}$ and $1000\;\mathrm{\mu m}$, respectively (Fig.~\ref{figure:YIG_SMR_Introduction}(c)). We measure the magnetotransport at temperatures $T$ between 5 K and 300 K in a liquid He magnet cryostat system in magnetic fields $H$ of up to $7\;\mathrm{T}$. The magnetoresistance was studied by applying a constant dc bias current density $J_\mathrm{q}$ in the range of $0.1\;\mathrm{kA/cm^2}$ to $500\;\mathrm{kA/cm^2}$ along the Hall bar while recording the longitudinal $V_\mathrm{long}$ and transverse $V_\mathrm{trans}$ voltage signals (Fig.~\ref{figure:YIG_SMR_Introduction}(c)) of the Hall bar as a function of the external magnetic field magnitude $H$ or orientation $\mathbf{h}$. For the further evaluation we calculated the longitudinal $\rho_\mathrm{long}=V_\mathrm{long}(J_\mathrm{q}l)^{-1}$ and transverse resistivity $\rho_\mathrm{trans}=V_\mathrm{trans}(J_\mathrm{q}w)^{-1}$, where $l=600\;\mathrm{\mu m}$ is the separation between the longitudinal voltage contacts and $w=80\;\mathrm{\mu m}$ the width of the Hall bar. Within the range of charge current densities $J_\mathrm{q}$ quoted above, the resistivities were independent of $J_\mathrm{q}$. We note that a reference sample consisting of a single, laser-MBE grown YIG ($30\;\mathrm{nm}$) layer on a GGG substrate was found to be electrically insulating within our experimental limits. This gives a lower limit for the resistivity of our YIG layers $\rho_\mathrm{YIG}>6\times10^{2}\;\mathrm{\Omega m}$, which is in agreement with the reported resistivity values for bulk YIG~\cite{sirdeshmukh_dielectric_1998} ($\rho_\mathrm{YIG}=6\times10^{7}\;\mathrm{\Omega m}$).

In a first set of experiments we verified the existence of a magnetoresistance effect in our hybrids~\cite{weiler_local_2012,Nakayama2012,huang_transport_2012,[{Shortly before submission of the manuscript, we became aware of this publication on the arXiv: }]vlietstra_spin-hall_2013} by recording the resistance evolution as a function of the applied external magnetic field strength ($R(H)$ measurements).
\begin{figure}[t,b]
  \includegraphics[width=85mm]{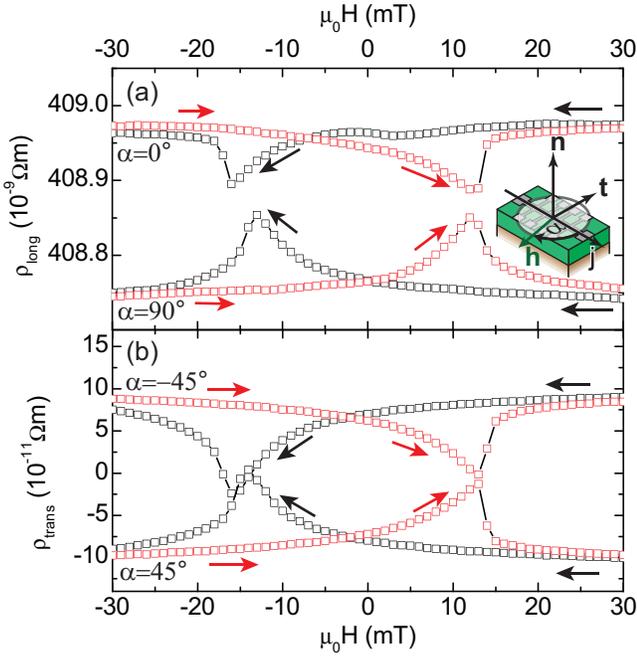}\\
  \caption{(Color online) $R(H)$ measurements conducted on a laser-MBE grown YIG($54\;\mathrm{nm}$)/Pt($7\;\mathrm{nm}$) hybrid structure at $T=300\;\mathrm{K}$ grown on a (111)-oriented GGG substrate. (a) Longitudinal resistivity $\rho_\mathrm{long}$ as a function of the applied in-plane magnetic field $H$ with $\mathbf{H}$ oriented parallel ($\alpha=0^\circ$) and perpendicular ($\alpha=90^\circ$) to the current direction $\mathbf{j}$. A clear resistance hysteresis behavior is visible in both orientations. The inset defines the magnetic field orientation angle $\alpha$ with respect to $\mathbf{j}$. (The arrow indicates the positive angle direction.) (b) Hysteresis curve of the transverse resistivity for the external field applied at $\alpha=-45^\circ$ and $\alpha=45^\circ$. In both panels black symbols represent data taken during the $H$-field down-sweep, and red symbols represent data from the $H$-field up-sweep, as illustrated by the black and red arrows in the graphs.}
  \label{RHPtYIG}
\end{figure}
Figure~\ref{RHPtYIG}(a) and (b) show the $R(H)$ results obtained at $T=300\;\mathrm{K}$ from YIG($54\;\mathrm{nm}$)/Pt($7\;\mathrm{nm}$) bilayer grown on a (111)-oriented GGG substrate via laser-MBE. For the external $H$-field oriented parallel to the current direction $\mathbf{j}$ ($\alpha=0^\circ$) we observe characteristic hysteretic changes of $\rho_\mathrm{long}$ around magnetic field values, that are larger than the coercive field values (cf.~Fig.~\ref{LaserMBEYIG_Pt}(d)). We attribute these differences to the Hall bar structuring process of the YIG/Pt hybrids. Upon changing the in-plane field orientation such that $\mathbf{H}$ is oriented perpendicular to $\mathbf{j}$ ($\alpha=90^\circ$), the hysteretic behavior is inverted. Moreover, the resistivity values for $H\gg H_\mathrm{c}$ are clearly different for $\alpha=0^\circ$ and $\alpha=90^\circ$. This dependence of the magnetoresistance on the external field orientation is expected from our theoretical model of the SMR (cf. Eq.(\ref{equ:YIG_SMR_rho_long})) and reflects the evolution of the magnetization orientation $\mathbf{m}=\left(m_\mathrm{j},m_\mathrm{t},m_\mathrm{n}\right)$ as a function of the external magnetic field magnitude in close analogy to the AMR effect. A similar behavior is expected for the transverse resistivity $\rho_\mathrm{trans}(H)$. From Eq.(\ref{equ:YIG_SMR_rho_trans}) the characteristic inversion should be most prominent for magnetic field orientations $\alpha=-45^\circ$ and $\alpha=45^\circ$. Our measurements in Fig.~\ref{RHPtYIG}(b) clearly confirm these expectations.
Unfortunately, it is not straightforward to distinguish directly between the polycrystalline AMR (Eqs.(\ref{equ:Poly_AMR_rho_long}), (\ref{equ:Poly_AMR_rho_trans})) and the SMR (Eqs.(\ref{equ:YIG_SMR_rho_long}), (\ref{equ:YIG_SMR_rho_trans})) effect from single $R(H)$ measurements. Moreover, a quantitative evaluation of the MR effect from these type of measurements is difficult, since magnetic domain formation around $H_\mathrm{c}$ influences the observed MR signal, while domain configurations are not taken into account in Eqs.(\ref{equ:YIG_SMR_rho_long}), (\ref{equ:YIG_SMR_rho_trans}) or Eqs.(\ref{equ:Poly_AMR_rho_long}), (\ref{equ:Poly_AMR_rho_trans}).

\begin{figure*}[h,b,t]
  \includegraphics[width=170mm]{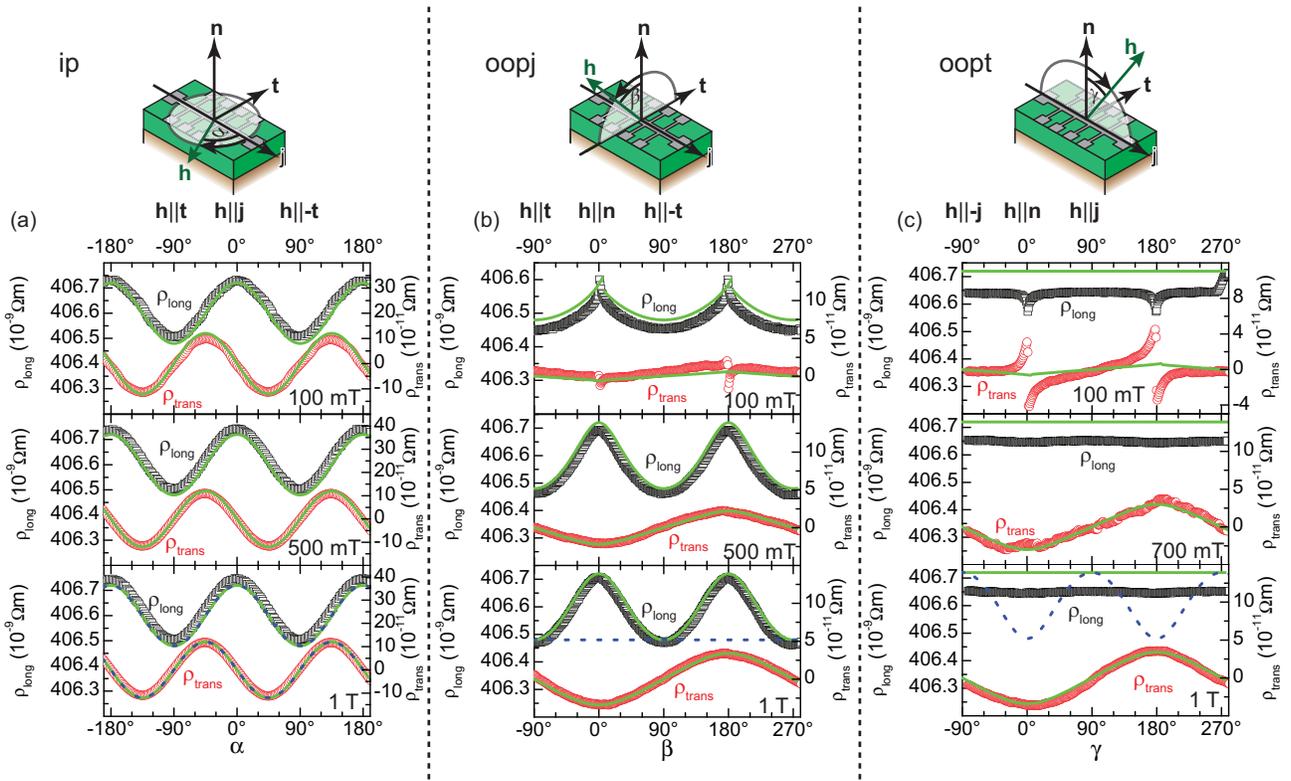}\\
  \caption[Spin reflection magnetoresistance for YIG/Pt heterostructures evaluated in various ADMR rotation planes]{(Color online) ADMR results obtained from a laser-MBE grown YIG ($54\;\mathrm{nm}$)/Pt ($7\;\mathrm{nm}$) hybrid structure on a (111)-oriented GGG substrate at $T=300\;\mathrm{K}$. (a) Evolution of $\rho_\mathrm{long}$ (black squares) and $\rho_\mathrm{trans}$ (red circles) as a function of $\mathbf{h}$-orientation for an in-plane (ip) rotation of the external magnetic field at $\mu_0 H_\mathrm{meas}=1\;\mathrm{T}$, $\mu_0 H_\mathrm{meas}=500\;\mathrm{mT}$, and $\mu_0 H_\mathrm{meas}=100\;\mathrm{mT}$. The ip angle $\alpha$ is defined in the illustration above the data panels (the arrow indicates positive direction). (b) Dependence of $\rho_\mathrm{long}$ and $\rho_\mathrm{trans}$ on the oopj magnetic field angle $\beta$ at $\mu_0 H_\mathrm{meas}=1\;\mathrm{T}$, $\mu_0 H_\mathrm{meas}=500\;\mathrm{mT}$, and $\mu_0 H_\mathrm{meas}=100\;\mathrm{mT}$. The positive angle $\beta$ is illustrated in the drawing above the data plots. (c) Angular evolution of $\rho_\mathrm{long}$ and $\rho_\mathrm{trans}$ for the oopt geometry at $\mu_0 H_\mathrm{meas}=1\;\mathrm{T}$, $\mu_0 H_\mathrm{meas}=700\;\mathrm{mT}$, and $\mu_0 H_\mathrm{meas}=100\;\mathrm{mT}$. The oopt angle $\gamma$ is shown in the sketch above the experimental data. In panels (a), (b), (c), the green lines represent fits to the data using the SMR model defined by Eqs.(\ref{equ:YIG_SMR_rho_long}), (\ref{equ:YIG_SMR_rho_trans}) with parameters of Table~\ref{table:YIG_SMR_SimResults}, while the dashed blue lines represent fits to the data using the AMR expression defined by Eqs.(\ref{equ:Poly_AMR_rho_long}), (\ref{equ:Poly_AMR_rho_trans}). Interestingly, the angular dependence of $\rho_\mathrm{long}$ vanishes for the oopt geometry, which is not consistent with the AMR of a polycrystalline FM, but indeed is consistent with the behavior expected for the SMR effect.}
  \label{figure:YIG_SMR_ADMR}
\end{figure*}

ADMR measurements are not hampered by domain configuration issues. In an ADMR measurement an external magnetic field of constant magnitude $H_\mathrm{meas}\gg H_\mathrm{c}$ is rotated in a given plane with respect to the sample, while $\rho_\mathrm{long}$ and $\rho_\mathrm{trans}$ are recorded as a function of this external magnetic field orientation. We here performed ADMR experiments in 3 rotation planes: ip, oopj, oopt. Herby, ip stands for a rotation of $\mathbf{H}$ around $\mathbf{n}$ (angle $\alpha$), oopj for a rotation of $\mathbf{H}$ around $\mathbf{j}$ (angle $\beta$), and oopt for a rotation of $\mathbf{H}$ around $\mathbf{t}$ (angle $\gamma$). For the definitions of $\alpha$, $\beta$, and $\gamma$ see the illustrations in Fig.~\ref{figure:YIG_SMR_ADMR}(a),(b), and (c), respectively, in which the field orientation $\mathbf{h}=\mathbf{H}/H$ is also shown. Prior to the angular variation we initialize the magnetization by applying $\mu_0 H_\mathrm{init}=3\;\mathrm{T}$ along $\alpha=0^\circ$ (ip), $\beta=\gamma=-90^\circ$ (oopj, oopt). In our measurements we choose $H_\mathrm{meas}$ to be at least one order of magnitude larger than the coercive field of YIG, such that the FMI is in a single domain state.

We begin the evaluation of the SMR with the analysis of the ADMR results obtained from a laser-MBE grown YIG ($54\;\mathrm{nm}$)/Pt ($7\;\mathrm{nm}$) hybrid structure at $T=300\;\mathrm{K}$. The experimental results are summarized in Fig.~\ref{figure:YIG_SMR_ADMR}(a)-(c).
For the ADMR experiments with an ip magnetic field rotation at $\mu_0 H_\mathrm{meas}=1\;\mathrm{T}$, $\mu_0 H_\mathrm{meas}=500\;\mathrm{mT}$, and $\mu_0 H_\mathrm{meas}=100\;\mathrm{mT}$, depicted in Fig.~\ref{figure:YIG_SMR_ADMR}(a), we observe an angular dependence with a period of $180^\circ$ and a maximum in $\rho_\mathrm{long}$ for $\mathbf{h}$ parallel ($\alpha=0^\circ$) or antiparallel ($\alpha=180^\circ$) to $\mathbf{j}$ and a minimum for $\mathbf{h}$ parallel ($\alpha=-90^\circ$) or antiparallel ($\alpha=90^\circ$) to $\mathbf{t}$ for every magnitude $H_\mathrm{meas}$. Due to the small in-plane magnetic anisotropy of the YIG layer its magnetization is oriented parallel to the external magnetic field in good approximation for all $\alpha$. This allows to directly use $\alpha$ to determine the magnetization orientation. Thus $\rho_\mathrm{long}$ is expected to follow a $\cos^2 \alpha$ dependence in accordance to Eq.(\ref{equ:YIG_SMR_rho_long}), which is nicely reproduced by our ADMR data for all magnetic fields (lines in Fig.~\ref{figure:YIG_SMR_ADMR}(a)). In addition, the transverse resistance exhibits a $\cos\alpha\sin\alpha$ dependence. We observe a maximum in $\rho_\mathrm{trans}$ at $\alpha=-45^\circ$ and $\alpha=135^\circ$, while the minimum is located at $\alpha=45^\circ$ and $\alpha=-135^\circ$. For both $\rho_\mathrm{long}$ and $\rho_\mathrm{trans}$ the amplitude of the angular dependence is not influenced by the external magnetic field strength, and $\rho_1=-\rho_3$ as expected.

\begin{figure}[h,b,t]
  \includegraphics[width=85mm]{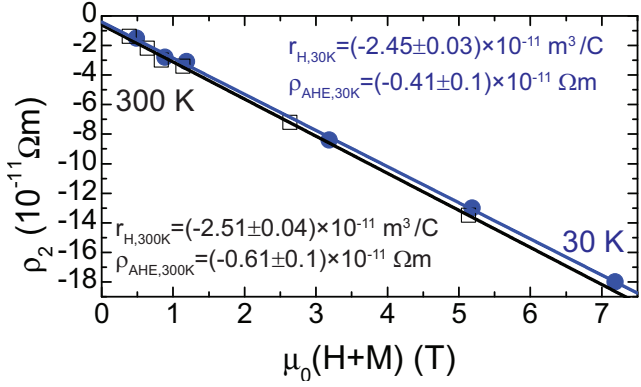}\\
  \caption[Extracted field dependence of $\rho_2$]{(Color online) Magnetic inductance dependence of $\rho_2$ for a laser-MBE grown YIG ($54\;\mathrm{nm}$)/Pt ($7\;\mathrm{nm}$) hybrid structure extracted from the fit to the ADMR data at $T=300\;\mathrm{K}$ (black open squares) and $T=30\;\mathrm{K}$ (blue closed circles). The black and blue lines represent linear fits to determine the Hall coefficients $r_\mathrm{Hall}$ and the AHE contributions $\rho_\mathrm{AHE}$ for $T=300\;\mathrm{K}$ and $T=30\;\mathrm{K}$, respectively.}
  \label{figure:YIG_SMR_AHE}
\end{figure}

In case of the oopj rotation plane (Fig.~\ref{figure:YIG_SMR_ADMR}(b)) we observe maxima in $\rho_\mathrm{long}$ at $\mu_0 H_\mathrm{meas}=1\;\mathrm{T}$ and $\mu_0 H_\mathrm{meas}=500\;\mathrm{mT}$ located at $\beta=0^\circ$ ($\mathbf{h}\parallel\mathbf{n}$) and $\beta=180^\circ$ ($\mathbf{h}\parallel-\mathbf{n}$). The minima in $\rho_\mathrm{long}$ occur at $\beta=-90^\circ$ ($\mathbf{h}\parallel\mathbf{t}$) and $\beta=90^\circ$ ($\mathbf{h}\parallel-\mathbf{t}$). At $\mu_0 H_\mathrm{meas}=1\;\mathrm{T}$ the longitudinal resistivity again exhibits a $\cos^2 \beta$ dependence. Upon reducing $H_\mathrm{meas}$ the shape anisotropy increasingly influences the orientation of the magnetization in YIG, which results in deviations from the $\cos^2 \beta$ dependence.~\cite{limmer_angle-dependent_2006} For the transverse resistivity we now observe a completely different angular dependence with a period of $360^\circ$. $\rho_\mathrm{trans}$ has a minimum at $\beta=0^\circ$ ($\mathbf{h}\parallel\mathbf{n}$) and a maximum at $\beta=180^\circ$ ($\mathbf{h}\parallel-\mathbf{n}$). The amplitude of this $\cos \beta$ dependence of $\rho_\mathrm{trans}$ strongly dependens on the external magnetic field strength (cf.~Fig.~\ref{figure:YIG_SMR_AHE}). This is the signature of the ordinary Hall effect of the Pt layer. The abrupt changes at $\mu_0 H_\mathrm{meas}=100\;\mathrm{mT}$, visible in $\rho_\mathrm{long}$ and $\rho_\mathrm{trans}$, originate from the in-plane reorientation of the magnetization forced by the demagnetization energy, if the field is rotated near $\beta=0^\circ$ or $180^\circ$.

Interestingly, the angular dependence of $\rho_\mathrm{long}(\gamma)$ vanishes for the oopt rotation plane (Fig.~\ref{figure:YIG_SMR_ADMR}(c)). This is in contrast to the expected angular dependence of an AMR of a polycrystalline FM (cf. Eq.(\ref{equ:Poly_AMR_rho_long})), but is fully consistent with our SMR model (cf. Eq.(\ref{equ:YIG_SMR_rho_long})). Thus, we can exclude an AMR effect as the source of the observed MR. For $\rho_\mathrm{trans}$ we again find a $\cos(\gamma)$ angular dependence, which dominantly stems from the ordinary Hall effect in Pt. The abrupt changes in $\rho_\mathrm{long}$ and $\rho_\mathrm{trans}$ visible at $\mu_0 H_\mathrm{meas}=100\;\mathrm{mT}$ are explained by the abrupt in-plane reorientation of the magnetization, when the field orientation is rotated near $\gamma=0^\circ$ or $\gamma=180^\circ$.

Huang \textit{et al.} also observed a MR in YIG/Pt bilayers~\cite{huang_transport_2012} and attributed it to an AMR effect originating from an induced magnetism (static magnetic proximity effect) in the Pt layer. As discussed in the context of Fig.~\ref{figure:YIG_SMR_ADMR}, our results suggest that systematic measurements as a function of $\mathbf{M}$ orientation in out-of-plane geometries allow to distinguish between SMR and AMR. The angular dependence we observe in our samples (Fig.~\ref{figure:YIG_SMR_ADMR}(c)) can be consistently explained by the SMR model presented in Sect.~\ref{Sect_Theory_SMR}, while conventional AMR can be ruled out as an explanation for the observed magnetoresistance.

For a quantitative analysis, we now employ the simulation technique successfully used for ADMR in (Ga,Mn)As~\cite{bihler_ga_1xmn_xas/piezoelectric_2008,limmer_advanced_2008,limmer_angle-dependent_2006} and Heusler compounds~\cite{muduli_antisymmetric_2005}. First we choose a starting value for the anisotropy constant $B_\mathbf{n}$ and determine the magnetization direction $\mathbf{m}$ for every magnetic field orientation by numerically minimizing the free enthalpy given by Eq.(\ref{equ:YIG_SMR_Free_energy}). We then calculate $\rho_\mathrm{long}$ and $\rho_\mathrm{trans}$ using $\mathbf{m}$ and a set of $\rho_i$ parameters via Eqs.(\ref{equ:YIG_SMR_rho_long}), (\ref{equ:YIG_SMR_rho_trans}). An iterative optimization process of $B_\mathbf{n}$ and the $\rho_i$ parameters is carried out, until we achieve a satisfactory agreement ($\sum \chi^2\leq10^{-6}$) between experiment and simulation for all rotation planes and $H_\mathrm{meas}$ magnitudes with a single set of $B_\mathbf{n}$ and $\rho_i$. For the simulation, all $\rho_i$ parameters except $\rho_2$ were taken to be independent of the external magnetic field strength. The simulation curves obtained in this way are depicted as green lines in Fig.~\ref{figure:YIG_SMR_ADMR}(a)-(c) and reproduce the experimental data very well. The respective $\rho_i$ parameters are summarized in Table~\ref{table:YIG_SMR_SimResults} and we obtained $B_\mathbf{n}=75\;\mathrm{mT}$ from the simulation.
\begin{table*}[t,b]
\begin{center}
\begin{tabular}{|l|c|c|c|c|c|}
  \hline
  sample & $T$ [K] & $\rho_0$ [$10^{-9}\Omega$m] & $\rho_1/\rho_0$ & $\rho_3/\rho_0$ & $\rho_2 (1\;\mathrm{T})$ [$10^{-11}\Omega$m]\\
  \hline 
  GGG/YIG(54)/Pt(7) & 300 & 406.5 & $-5.9\times10^{-4}$ & $5.9\times10^{-4}$ & $-3.4$\\
  GGG/YIG(54)/Pt(7) & 30 & 283.5 & $-4.7\times10^{-4}$ & $4.7\times10^{-4}$ & $-3.1$\\
  GGG/YIG(54)/Pt(7) & 5 & 208.7 & $-3.9\times10^{-4}$ & $3.9\times10^{-4}$ & $-2.2$\\
  \hline 
  GGG/YIG(61)/Pt(1.1) & 300 & 1895.9 & $-3.6\times10^{-4}$ & $3.6\times10^{-4}$ & $-2.5$\\
  \hline 
  GGG/YIG(55)/Pt(1.2) & 300 & 1857.3 & $-3.9\times10^{-4}$ & $3.9\times10^{-4}$ & $-1.5$\\
  \hline 
  GGG/YIG(57)/Pt(1.3) & 300 & 1089.9 & $-3.5\times10^{-4}$ & $3.5\times10^{-4}$ & $-18.0$\\
  \hline 
  GGG/YIG(29)/Pt(1.8) & 300 & 487.4 & $-1.5\times10^{-3}$ & $1.5\times10^{-3}$ & $-2.5$\\
  \hline 
  GGG/YIG(58)/Pt(2.2) & 300 & 761.7 & $-1.2\times10^{-3}$ & $1.2\times10^{-3}$ & $-3.5$\\
  \hline 
  GGG/YIG(69)/Pt(2.7) & 300 & 453.6 & $-1.5\times10^{-3}$ & $1.5\times10^{-3}$ & $-4.8$\\
  \hline 
  GGG/YIG(53)/Pt(2.5) & 300 & 719.0 & $-1.6\times10^{-3}$ & $1.6\times10^{-3}$ & $-5.8$\\
  \hline 
  GGG/YIG(46)/Pt(3.5) & 300 & 306.6 & $-9.4\times10^{-4}$ & $9.4\times10^{-4}$ & $-3.0$\\
  \hline 
  GGG/YIG(65)/Pt(6.6) & 300 & 582.6 & $-6.4\times10^{-4}$ & $6.4\times10^{-4}$ & $-7.8$\\
  \hline 
  GGG/YIG(50)/Pt(7) & 300 & 409.4 & $-6.1\times10^{-4}$ & $6.1\times10^{-4}$ & $-3.4$\\
  \hline 
  GGG/YIG(53)/Pt(8.5) & 300 & 348.3 & $-6.4\times10^{-4}$ & $6.4\times10^{-4}$ & $-3.5$\\
  \hline 
  GGG/YIG(61)/Pt(11.1) & 300 & 334.5 & $-4.4\times10^{-4}$ & $4.4\times10^{-4}$ & $-3.0$\\
  \hline 
  GGG/YIG(52)/Pt(16.9) & 300 & 339.2 & $-3.2\times10^{-4}$ & $3.2\times10^{-4}$ & $-2.9$\\
  \hline 
  GGG/YIG(55)/Pt(17.2) & 300 & 331.7 & $-3.0\times10^{-4}$ & $3.0\times10^{-4}$ & $-2.3$\\
  \hline
  \hline 
  YAG/YIG(45)/Pt(1.9) & 300 & 1331.3 & $-6.8\times10^{-4}$ & $6.8\times10^{-4}$ & $-3.5$\\
  \hline 
  YAG/YIG(60)/Pt(2.5) & 300 & 358.4 & $-1.1\times10^{-3}$ & $1.1\times10^{-3}$ & $-2.8$\\
  \hline 
  YAG/YIG(64)/Pt(3) & 300 & 622.2 & $-1.4\times10^{-3}$ & $1.4\times10^{-3}$ & $-3.2$\\
  \hline 
  YAG/YIG(50)/Pt(3) & 300  & 539.3 & $-1.2\times10^{-3}$ & $1.2\times10^{-3}$ & $-4.0$\\
  \hline 
  YAG/YIG(63)/Pt(6.5) & 300  & 412.0 & $-7.9\times10^{-4}$ & $7.9\times10^{-4}$ & $-3.4$\\
  \hline 
  YAG/YIG(59)/Pt(6.8) & 300  & 487.7 & $-9.5\times10^{-4}$ & $9.5\times10^{-4}$ & $-3.5$\\
  \hline 
  YAG/YIG(60)/Pt(9.7) & 300  & 429.0 & $-5.7\times10^{-4}$ & $5.7\times10^{-4}$ & $-3.9$\\
  \hline 
  YAG/YIG(60)/Pt(12.8) & 300  & 434.9 & $-4.3\times10^{-4}$ & $4.3\times10^{-4}$ & $-3.5$\\
  \hline 
  YAG/YIG(61)/Pt(19.5) & 300  & 361.3 & $-2.8\times10^{-4}$ & $2.8\times10^{-4}$ & $-2.9$\\
  \hline 
  YAG/YIG(46)/Pt(21.7) & 300  & 358.7 & $-2.3\times10^{-4}$ & $2.3\times10^{-4}$ & $-2.7$\\
  \hline
  \hline 
  GGG/YIG(15)/Au(8)/Pt(7)  & 300 & 143.0 & $-2.4\times10^{-4}$ & $2.4\times10^{-4}$ & $-11$ \\
  \hline 
  GGG/YIG(34)/Cu(9)/Pt(6) & 300 & 111.0 & $-0.9\times10^{-4}$ & $0.9\times10^{-4}$ & $-8$\\
  \hline
  \hline
  MgO/Fe$_3$O$_4$(20)/Pt(7) & 5 & 202.6 & $-2.1\times10^{-4}$ & $2.1\times10^{-4}$ & $-4.2$\\
  \hline
  \hline
  MgAl$_2$O$_4$/NiFe$_2$O$_4$(620)/Pt(10)& 300 & 242.0 & $-2.7\times10^{-4}$ & $2.7\times10^{-4}$ & $-2.2$\\
  \hline
\end{tabular}

\caption{The $\rho_i$ parameters obtained from fits to the experimental data for FMI/Pt hybrids and FMI/NM/Pt hybrids investigated in this work. The numbers in parentheses give the thickness of the respective layers in nm. All YIG based heterostructures in this table were fabricated via laser-MBE. The $\rho_i$ parameters for the Fe$_3$O$_4$/Pt hybrid structure have been extracted at $\mu_0H=1\;\mathrm{T}$.}
\label{table:YIG_SMR_SimResults}
\end{center}
\end{table*}
The $\rho_i$ parameters extracted from our simulation show $\rho_1=-\rho_3$. This corroborates the prediction based on the qualitative description of the SMR effect.

The MR of the laser-MBE grown YIG/Pt heterostructure only changes slightly with temperature. At $T=5\;\mathrm{K}$, $\rho_1/\rho_0$ is reduced by 34\% as compared to $T=300\;\mathrm{K}$. This decrease can be caused by a temperature dependence of the spin mixing conductance~\cite{czeschka_scaling_2011}, the spin diffusion length, and the spin Hall angle. However, more experiments are required to further clarify this issue.
Moreover, $\rho_1/\rho_0$ is independent of the current density in the sample, such that thermal gradients generated via Joule heating can be excluded as a source of the MR effect.

\begin{figure}[h,b,t]
  \includegraphics[width=85mm]{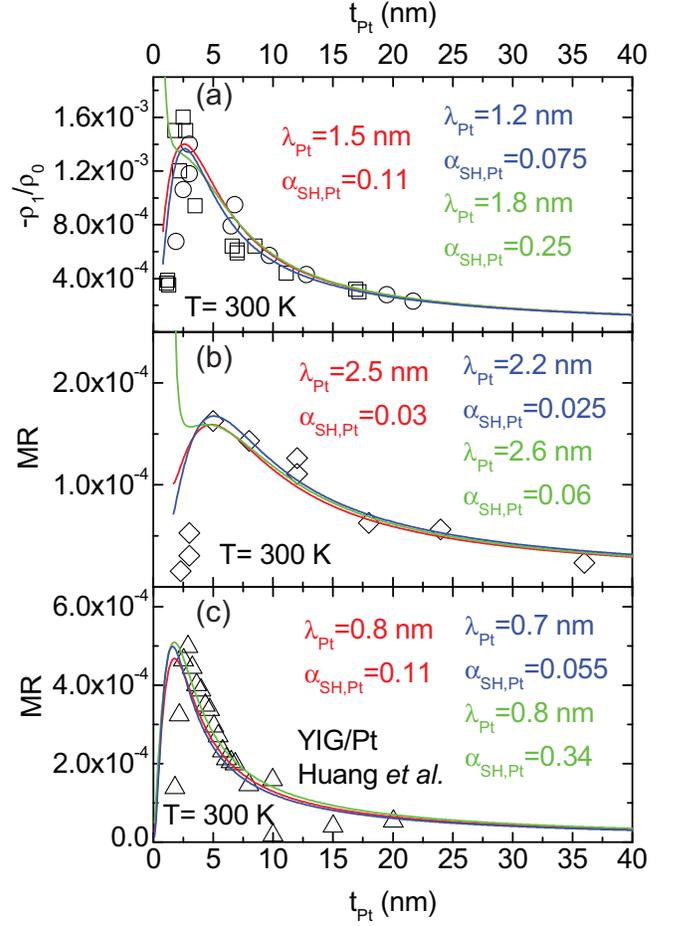}\\
  \caption[Pt thickness dependence of the SMR signal]{(Color online) Evolution of the SMR effect as a function of Pt thickness $t_\mathrm{Pt}$. (a) $t_\mathrm{Pt}$ dependence of $\rho_1/\rho_0$ determined from different laser-MBE grown YIG/Pt ($[t_\mathrm{Pt}]\;\mathrm{nm}$) samples (open symbols, squares for samples on GGG substrate, circles for samples on YAG substrates) from our fits to the ADMR data. (b) $t_\mathrm{Pt}$ dependence of the SMR signal for liquid phase epitaxy grown YIG and sputter deposited Pt heterostrucutures extracted from $R(H)$ experiments~\cite{Nakayama2012}. (c) Dependence of the MR on $t_\mathrm{Pt}$ determined by Huang \textit{et al.}~\cite{huang_transport_2012} for YIG/Pt hybrids. The lines in each panel represent a simulation based on Eq.(\ref{equ:SMR_Quan_Ratio}) for $G_r=4\times10^{14}\;\mathrm{\Omega^{-1}m^{-2}}$ (red), $G_r=4\times10^{13}\;\mathrm{\Omega^{-1}m^{-2}}$ (green), and $G_r=4\times10^{15}\;\mathrm{\Omega^{-1}m^{-2}}$ (blue). From this simulation we extract the spin Hall angle $\alpha_\mathrm{SH}$ and the spin diffusion length $\lambda_\mathrm{Pt}$ in Pt quoted in the respective panels (same color coding).}
  \label{figure:YIG_SMR_Pt_thickness}
\end{figure}

From the quantitative description of the SMR effect (Eq.(\ref{equ:SMR_Quan_Ratio})) it is evident that the SMR sensitively depends on the Pt layer thickness. We have thus used a set of samples with different Pt thicknesses and evaluated the SMR effect from ADMR experiments to extract the $\rho_i$ parameters in an iteration process as described above. The results of this procedure are compiled in Table~\ref{table:YIG_SMR_SimResults}. The ratio $-\rho_1/\rho_0$ is plotted as a function of the Pt thickness in Fig.~\ref{figure:YIG_SMR_Pt_thickness}(a).
We clearly observe a maximum of the SMR in our samples (black symbols) at a Pt thickness of around $3\;\mathrm{nm}$ with a ratio of $-\rho_1/\rho_0=1.6\times10^{-3}$. A simulation of the experimental data using Eq.(\ref{equ:SMR_Quan_Ratio}) allows us to extract the relevant material parameters from this film thickness dependence of the SMR. We however would like to emphasize that the parameter values thus obtained sensitively depend on $\rho_\mathrm{Pt}$ and the value used for $G_r$. In the analysis, we explicitly  took the film thickness dependence of $\rho_\mathrm{Pt}$ into account by fitting Eq.(\ref{equ:AppendixRho}) to the Pt thickness dependence of $\rho_0$ (see Fig.~\ref{figure:rho_Pt_thickness}(a)) in the simulation. Note also that for values of $t_\mathrm{Pt}$ that are smaller than the surface roughness $h$, Eq.(\ref{equ:AppendixRho}) is no longer applicable, which puts a lower limit on the range of Pt thicknesses considered in the simulation. We note that $l_\infty$ is the charge transport mean free path for an infinitely thick film (see Appendix~\ref{Appendix}); for a finite thickness the charge transport mean free path $\ell$ decreases with decreasing thickness.~\cite{fischer_mean_1980}. When surface roughness limits transport ($t_\mathrm{Pt}\leq l_\infty$) our SMR model based on the spin diffusion equation is no longer applicable. We can model such a regime by introducing a spin diffusion length or spin-flip relaxation time that depends on thickness of the metal layer.~\cite{valenzuela_direct_2006,erekhinsky_surface_2010,poli_spin-flip_2006,vila_evolution_2007} A transport theory based on Boltzmann equations is better suited to treat the thickness dependence of the SMR in the presence of significant surface roughness scattering. However, such a more elaborate theoretical description is way beyond the scope of the present paper. In particular, the solutions of a Boltzmann-type spin transport approach most likely will not be analytical anymore. In contrast, the simple spin diffusion model we use here does yield analytical expressions, which allows to pinpoint the microscopic nature of the physical mechanisms. Realizing the limitations of our model, we focus on samples with $t_\mathrm{Pt}\geq 5\;\mathrm{nm}$ for the parameter optimization, since in this limit, the spin diffusion approach should yield a reasonable description of the spin transport processes. To capture the impact of $G_r$ we use three different typical values to show the influence of this parameter on the extracted $\alpha_\mathrm{SH}$ and $\lambda_\mathrm{Pt}$. For $G_r=4\times10^{14}\;\mathrm{\Omega^{-1}m^{-2}}$ (cf.~Jia \textit{et al.}~\cite{jia_spin_2011}) we obtain a satisfactory agreement between simulation and experiment (red line in Fig. ~\ref{figure:YIG_SMR_Pt_thickness}(a)) for $\alpha_\mathrm{SH}=0.11$, $\lambda_\mathrm{Pt}=1.5\;\mathrm{nm}$. For a lower value $G_r=4\times10^{13}\;\mathrm{\Omega^{-1}m^{-2}}$, $\alpha_\mathrm{SH}=0.25$, $\lambda_\mathrm{Pt}=1.8\;\mathrm{nm}$ yield the best agreement between experiment and simulation (green line in Fig.~\ref{figure:YIG_SMR_Pt_thickness}(a)). By choosing a higher value $G_r=4\times10^{15}\;\mathrm{\Omega^{-1}m^{-2}}$, we achieve for $\alpha_\mathrm{SH}=0.075$, $\lambda_\mathrm{Pt}=1.2\;\mathrm{nm}$ the lowest deviation between experiment and simulation (blue line in Fig.~\ref{figure:YIG_SMR_Pt_thickness}(a)). However, the simulated SMR thickness dependence for $G_r=4\times10^{13}\;\mathrm{\Omega^{-1}m^{-2}}$ diverges for small $t_\mathrm{Pt}$, while the simulation for $G_r=4\times10^{14}\;\mathrm{\Omega^{-1}m^{-2}}$ and $G_r=4\times10^{15}\;\mathrm{\Omega^{-1}m^{-2}}$ nicely reproduces also the thickness dependence of the experimental data for small $t_\mathrm{Pt}$. Since the value $G_r=4\times10^{15}\;\mathrm{\Omega^{-1}m^{-2}}$ appears unreasonably high, when compared to the theoretical calculations by Jia \textit{et al.}.~\cite{jia_spin_2011}, we conclude that $G_r=4\times10^{14}\;\mathrm{\Omega^{-1}m^{-2}}$ is most appropriate for our samples. The modeling  for what we perceive to be reasonable $G_r$ values nicely reproduce the experimentally observed thickness dependence also in the limit of very thin Pt films ($t_\mathrm{Pt}\leq 5\;\mathrm{nm}$). While the validity of the diffusive model is questionable in this limit, the Boltzmann corrections appear to be small. Nevertheless, a more elaborate theoretical evaluation is needed to clarify this point.

In an independent set of experiments we investigated the Pt thickness dependence of the SMR in YIG/Pt heterostructures fabricated via liquid phase epitaxy (YIG) and sputtering (Pt) on (111)-oriented GGG substrates~\cite{Nakayama2012}. These samples were structured into Hall bars with $w=1000\;\mathrm{\mu m}$ and $l_\mathrm{HB}=2200\;\mathrm{\mu m}$. In this measurement series the longitudinal MR was determined from $R(H)$ curves for each sample. The results of this analysis are shown in Fig.~\ref{figure:YIG_SMR_Pt_thickness}(b). For these samples we observe a maximum of the MR for a Pt thickness of around $5\;\mathrm{nm}$ with a MR ratio of $1.4\times10^{-4}$. Following the same procedure as above, we included the thickness dependence of $\rho_\mathrm{Pt}$ via a fit of Eq.(\ref{equ:AppendixRho}) to the Pt thickness dependence of $\rho_0$ (see Fig.~\ref{figure:rho_Pt_thickness}(b)). For $G_r=4\times10^{14}\;\mathrm{\Omega^{-1}m^{-2}}$ (red line in Fig.~\ref{figure:YIG_SMR_Pt_thickness}(b)) we obtain $\alpha_\mathrm{SH}=0.03$, $\lambda_\mathrm{Pt}=2.5\;\mathrm{nm}$, with satisfactory agreement between simulation and experiment. For a smaller $G_r=4\times10^{13}\;\mathrm{\Omega^{-1}m^{-2}}$ (green line in Fig.~\ref{figure:YIG_SMR_Pt_thickness}(b)) the parameters change to $\alpha_\mathrm{SH}=0.06$, $\lambda_\mathrm{Pt}=2.6\;\mathrm{nm}$. A higher $G_r=4\times10^{15}\;\mathrm{\Omega^{-1}m^{-2}}$ (blue line in Fig.~\ref{figure:YIG_SMR_Pt_thickness}(b)) yields $\alpha_\mathrm{SH}=0.025$, $\lambda_\mathrm{Pt}=2.2\;\mathrm{nm}$. Again, the simulation for $G_r=4\times10^{14}\;\mathrm{\Omega^{-1}m^{-2}}$ reproduces the experimental thickness dependence very well and is consistent with the theoretical calculations of $G_r$ for YIG/noble metal interfaces.

In the publication of Huang \textit{et al.}~\cite{huang_transport_2012} the thickness dependence of the longitudinal MR in YIG/Pt heterostructures has also been investigated, the data taken from their publication are depicted in Fig.~\ref{figure:YIG_SMR_Pt_thickness}(c). Their MR thickness dependence has a maximum located also at a Pt thickness of $3\;\mathrm{nm}$ with a value of $5\times10^{-4}$. A simulation of their data set with Eq.(\ref{equ:SMR_Quan_Ratio}) using $\rho_\mathrm{Pt}=2.40\times 10^{-7}\;\mathrm{\Omega m}$ (we use an average Pt sheet resistivity as no data on the thickness dependence is given in this publication) yields $\lambda_\mathrm{Pt}=0.8\;\mathrm{nm}$ and $\alpha_\mathrm{SH}=0.11$ for $G_r=4\times10^{14}\;\mathrm{\Omega^{-1}m^{-2}}$, with an excellent agreement between data and simulation (red line in Fig.~\ref{figure:YIG_SMR_Pt_thickness}(c)). For $G_r=4\times10^{13}\;\mathrm{\Omega^{-1}m^{-2}}$ the parameters in the simulation change to $\lambda_\mathrm{Pt}=0.8\;\mathrm{nm}$ and $\alpha_\mathrm{SH}=0.34$ (green line in Fig.~\ref{figure:YIG_SMR_Pt_thickness}(c)). $G_r=4\times10^{15}\;\mathrm{\Omega^{-1}m^{-2}}$ (blue line in Fig.~\ref{figure:YIG_SMR_Pt_thickness}(c)) gives $\lambda_\mathrm{Pt}=0.7\;\mathrm{nm}$ and $\alpha_\mathrm{SH}=0.055$. As we now used a thickness independent $\rho_\mathrm{Pt}$ for the simulation in Fig.~\ref{figure:YIG_SMR_Pt_thickness}(c), the simulation does not diverge for small $t_\mathrm{Pt}$. While we anticipate that an inclusion of a thickness dependent $\rho_{Pt}$ will affect the values for $\alpha_\mathrm{SH}$ and $\lambda_\mathrm{Pt}$ to some extent, the good agreement between the experimental data of Huang \textit{et al.}~\cite{huang_transport_2012} and our SMR simulation suggest that the SMR effect also plays a crucial role in their experiment.

The spin diffusion lengths extracted from the three independent data sets in Fig.~\ref{figure:YIG_SMR_Pt_thickness} compare reasonably well. The extracted $\lambda_\mathrm{Pt}$ depends on the assumed $G_r$ and changes for our 3 evaluated $G_r$ values by $30\%$. While for a fixed $G_r$ the relative error for $\lambda_\mathrm{Pt}$ is less than $20\%$. Moreover, the spin diffusion length is comparable with the value of the charge transport mean free path in Pt (see Appendix~\ref{Appendix}). For $t_\mathrm{Pt}\leq 5\;\mathrm{nm}$ we find $\lambda_\mathrm{Pt}>\ell$, while $\lambda_\mathrm{Pt}<\ell$ is obtained for $t_\mathrm{Pt}> 5\;\mathrm{nm}$. A Boltzmann theory for the SMR at least at the level of the Fuchs-Sondheimer model will yield more insight into this problem. Also $\alpha_\mathrm{SH}$ is strongly correlated with $G_r$ and it is currently unclear whether size effects enhance or reduce the SMR effect magnitude, leading to an effective, thickness dependent $\alpha_\mathrm{SH}$. The parameter values extracted from our analysis thus should be considered with this caveat in mind. We attribute the differences in $\lambda_\mathrm{Pt}$ and $\alpha_\mathrm{SH}$ obtained for the different sets of samples to the different deposition techniques for the Pt layer. Nevertheless, the values obtained from Fig.~\ref{figure:YIG_SMR_Pt_thickness} deviate from those by Mosendz \textit{et al.}~\cite{mosendz_detection_2010} for Pt.
Calculations~\cite{gradhand_spin_2010} suggest that impurities can substantially change the magnitude of the spin Hall angle, while they only slightly alter the spin diffusion length in Pt. In one of our recent publications we investigated the spin pumping effect in various conductive ferromagnet/Pt heterostructures~\cite{czeschka_scaling_2011} and obtained a scaling relation, using the material constants quoted by Mosendz \textit{et al.}~\cite{mosendz_detection_2010} for Pt. We would like to note that the scaling relation also holds for the Pt parameter values we extracted from the SMR experiments in Fig.~\ref{figure:YIG_SMR_Pt_thickness}. Using $\alpha_\mathrm{SH}=0.11$ and $\lambda_\mathrm{Pt}=1.5\;\mathrm{nm}$ we obtain a spin mixing conductance in conductive ferromagnet/Pt bilayers of $G_r=4\times10^{14}\;\mathrm{\Omega^{-1}m^{-2}}$, corroborating the analysis of the SMR experiments above.

We now address the Hall effect in our laser-MBE grown YIG/Pt hybrids in more detail. The field dependence of $\rho_2$ for $T=300\;\mathrm{K}$ and $T=30\;\mathrm{K}$ extracted from the simulation of our ADMR data is shown in Fig.~\ref{figure:YIG_SMR_AHE} for the very same sample as shown in Fig.~\ref{figure:YIG_SMR_ADMR} (YIG ($54\;\mathrm{nm}$)/Pt ($7\;\mathrm{nm}$)). We here plot $\rho_2$ versus the magnetic inductance $\mu_0(H+M_\mathrm{YIG})$ to take into account the additional magnetic field in the Pt layer due to the magnetization of the YIG. The analysis is only conducted for $\mu_0H_\mathrm{meas}\geq 0.25\;\mathrm{T}$ as only then the magnetization of the YIG layer is saturated and oriented along the external magnetic field direction. For $T=300\;\mathrm{K}$ we use $M_\mathrm{YIG}=110\;\mathrm{kA/m}$ determined from SQUID magnetometry (Fig.~\ref{LaserMBEYIG_Pt}(d)). For $T=30\;\mathrm{K}$ we extrapolated the saturation magnetization from $300\;\mathrm{K}$ using the temperature dependence of bulk YIG~\cite{hansen_saturation_1974} and obtain $M_\mathrm{YIG}=150\;\mathrm{kA/m}$. A direct determination of the saturation magnetization via SQUID magnetometry at this temperature is not possible because of the paramagnetism of the GGG substrate. At both temperatures $\rho_2$ increases linearly with increasing magnetic field ($\mu_0(H+M_\mathrm{YIG})$), as one would expect for an OHE.~\cite{hall_new_1879} A linear fit to the data to extract the Hall coefficient of our Pt yields $r_\mathrm{Hall}\approx-2.5\times 10^{-11}\;\mathrm{m^3/C}$ for both temperatures. This value is close to $-2.1\times 10^{-11}\;\mathrm{m^3/C}$ reported in Ref.~\onlinecite{greig_hall_1972} for evaporated Pt films. Moreover, we find a non vanishing abscissa $\rho_\mathrm{AHE}=\left(-0.61\pm0.1\right)\times 10^{-11}\;\mathrm{\Omega m}$ at $T=300\;\mathrm{K}$ and $\rho_\mathrm{AHE}=\left(-0.41\pm0.1\right) \times 10^{-11}\;\mathrm{\Omega m}$ at $T=30\;\mathrm{K}$. This effect can be understood within our theoretical model~\cite{Chen_SMR_2013} of the SMR effect, which predicts an anomalous Hall-like SMR contribution due to the imaginary part of the spin mixing conductance (cf.~Eq(\ref{equ:AHE_SMR_Quan_Ratio})). From our experiment we obtain for $t_\mathrm{Pt}=7\;\mathrm{nm}$ a ratio $\rho_\mathrm{AHE}/\rho_0=1.5\times10^{-5}$ at $T=300\;\mathrm{K}$. Using the parameters $G_r=4\times10^{14}\;\mathrm{\Omega^{-1}m^{-2}}$, $\alpha_\mathrm{SH}=0.11$, $\lambda_\mathrm{Pt}=1.5\;\mathrm{nm}$ and Eq.(\ref{equ:AHE_SMR_Quan_Ratio}) we extract $G_i=1.1\times10^{13}\;\mathrm{\Omega^{-1}m^{-2}}$ from the experiment. This gives $G_i/G_r=0.03$ which nicely agrees with theoretical calculations~\cite{jia_spin_2011} ($G_i/G_r\approx1/20$). The quantitative agreement between theory and experiment for both magnetoresistance and Hall-type measurements confirms the existence of the SMR effect. Note that an induced ferromagnetism at the YIG/Pt interface~\cite{Gepraegs_YIG2012,huang_transport_2012,jaouen_ag-_2005,wilhelm_layer-resolved_2000,poulopoulos_structural_2003} might also contribute an AHE signal. However, our quantitative analysis of the Hall data, together with the MR data with out-of-plane magnetization have little room for such a static, magnetic proximity induced MR effect.

While the magnetoresistive behavior observed in our YIG/Pt samples is not consistent with an induced magnetization (static magnetic proximity effect) in the Pt close to the YIG/Pt interface, an experiment to rule out magnetic proximity as the origin of the SMR is desirable. In a recent publication,~\cite{Gepraegs_YIG2012} we employed XMCD measurements to detect element-specific magnetic moments at the Pt $L_3$-edge in the very same laser-MBE grown samples also used for SMR measurements presented here. The XMCD measurements showed that the induced magnetic proximity moment in Pt -- if present at all -- is small in our samples. However, the available XMCD data do not allow to exclude a static proximity effect in Pt, only an upper limit can be put on the induced Pt moment. Therefore, another set of experiments addressing the magnetic proximity effect in Pt is desirable.

\begin{figure*}[h,b,t]
  \includegraphics[width=170mm]{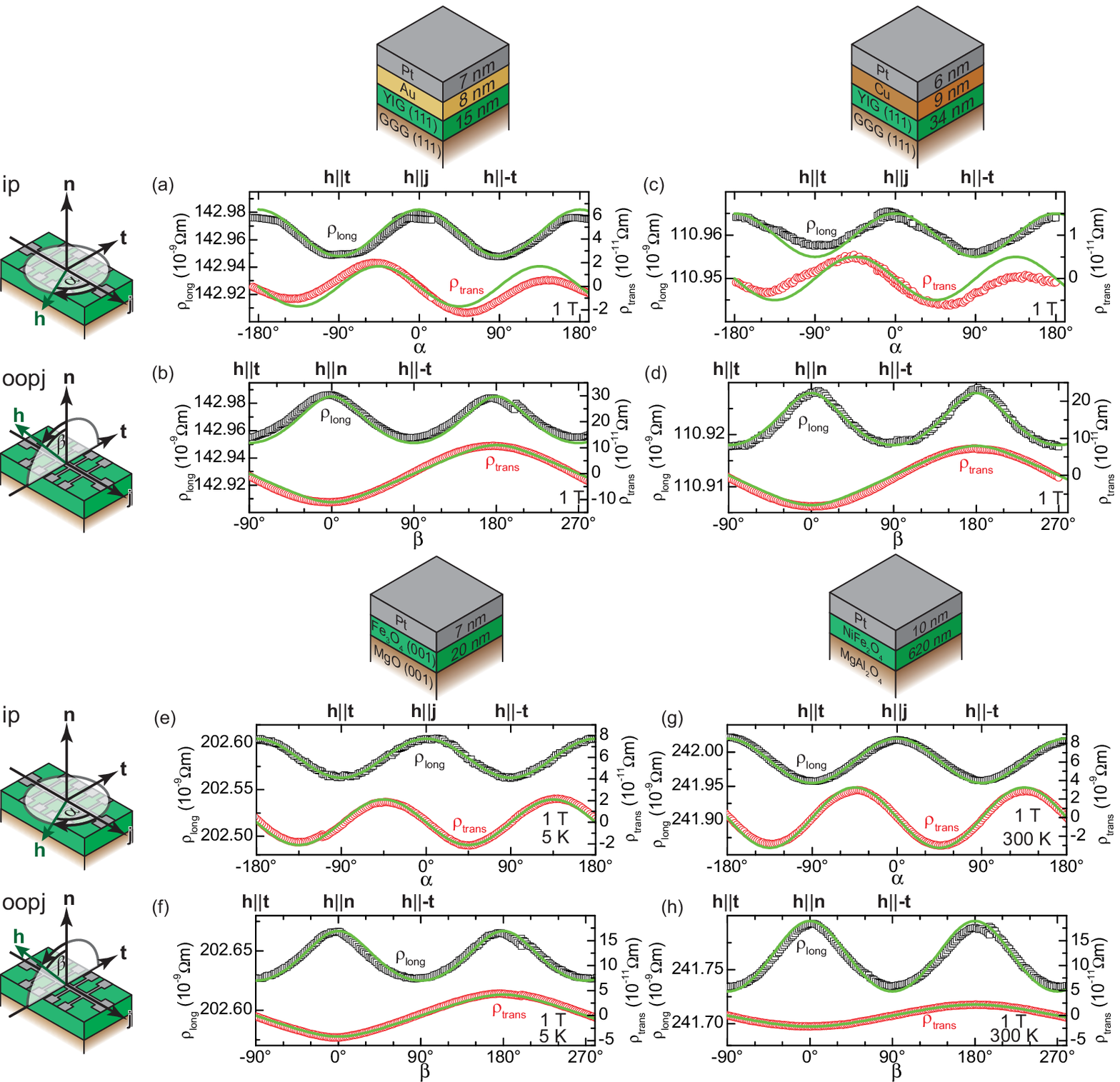}\\
  \caption[Spin reflection magnetoresistance for YIG/NM/Pt heterostructures]{(Color online) Magnetoresistance data (symbols) and corresponding SMR fits to Eqs.(\ref{equ:YIG_SMR_rho_long}), (\ref{equ:YIG_SMR_rho_trans}) (green lines) from laser-MBE grown YIG ($15\;\mathrm{nm}$)/Au ($8\;\mathrm{nm}$)/Pt ($7\;\mathrm{nm}$) (panels (a) and (b)), and laser-MBE grown YIG ($34\;\mathrm{nm}$)/Cu ($9\;\mathrm{nm}$)/Pt ($7\;\mathrm{nm}$) (panels (c) and (d)), Fe$_{3}$O$_{4}$ ($20\;\mathrm{nm}$)/Pt ($7\;\mathrm{nm}$) at $T=5\;\mathrm{K}$ and $\mu_0 H=1\;\mathrm{T}$ (panels (e) and (f)), NiFe$_{2}$O$_{4}$ ($620\;\mathrm{nm}$)/Pt ($10\;\mathrm{nm}$) (panels (g) and (h)) at $T=300\;\mathrm{K}$ and $\mu_0 H=1\;\mathrm{T}$. The oopt rotations (not shown here) exhibit no angular dependence in $\rho_\mathrm{long}$ for all 4 samples. The introduction of a second NM layer between YIG and Pt influences only the magnitude of the observed SMR. This clearly indicates that the SMR arises due to the spin current absorption at the YIG/NM interface and is not due to the MR behavior of an induced magnetic polarization in the Pt layer. The SMR signature also appears in other FMI/Pt hybrid structures and thus confirms the universality of the SMR effect.
  }
  \label{figure:YIG_SMR_Comparison}
\end{figure*}

Following the same line of argument as in our previous publication on the SMR,~\cite{Nakayama2012} we also investigated the impact of a NM layer between YIG and Pt on the SMR. In Fig.~\ref{figure:YIG_SMR_Comparison}(a)-(d) we show the ADMR signals of a laser-MBE grown YIG ($45\;\mathrm{nm}$)/Au ($7\;\mathrm{nm}$)/Pt ($7\;\mathrm{nm}$) and a laser-MBE grown YIG ($45\;\mathrm{nm}$)/Cu ($9\;\mathrm{nm}$)/Pt ($7\;\mathrm{nm}$) heterostructure. We note that the discrepancy between the green simulation curve and the experimental $\rho_\mathrm{trans}$ data in the ip rotation for both samples arises from a small out-of-plane external field component caused by a slightly tilted rotation plane ($2^\circ$). This experimental misalignment leads to a superimposed $\cos \alpha$ dependence. As the OHE and AHE contribution $\rho_2$ is comparable to $\rho_3$ in these samples, a slight tilting in the experiment from the ideal ip rotation plane already leads to a considerable deviation between experiment and simulation.
The ADMR signal (Fig.~\ref{figure:YIG_SMR_Comparison}(a)-(d)) from the laser-MBE grown YIG/NM/Pt samples is qualitatively identical to the ADMR from YIG/Pt (Fig.~\ref{figure:YIG_SMR_ADMR}). ADMR experiments in YIG (45 nm)/Au (7 nm) and YIG (45 nm)/Cu (7 nm) samples without Pt top layers did not reveal any variation of the resistance with $\mathbf{H}$ oriented in-plane within our experimental resolution. From these experiments, we obtain an upper limit of $-\rho_1/\rho_0\leq 5\times10^{-5}$ for YIG/NM samples without Pt (data not shown here). As mentioned in Sec.~\ref{Sect_Theory_SMR} the maximum $-\rho_1/\rho_0$ is strictly smaller than $\alpha_\mathrm{SH}^2$. Assuming $\alpha_\mathrm{SH}=0.0035$ for Au (Ref.~\onlinecite{mosendz_detection_2010}), one estimates $-\rho_1/\rho_0\leq 1.2\times10^{-5}$, which presently is below our experimental resolution. Thus, we are currently unable to resolve the SMR in FMI/Au hybrids. These results demonstrate that the MR effect observed in FMI/Pt and FMI/NM/Pt samples indeed is related to spin currents. In particular, a static magnetic proximity effect, i.e., an induced magnetic moment in the Pt layer, can be ruled out based on these experiments, since a magnetic proximity effect can not persist over 8 to 9 nm of Cu or Au.~\cite{bailey_pd_2012} For the SMR, only the conversion of a charge current into a spin current and back into a charge current via the SHE/ISHE in the Pt and the transport of spin current from the Pt to the YIG are necessary. Thus, the SMR persists even when a NM layer is inserted, provided that the spin current can propagate through this NM layer. The extracted quantitative data for all heterostructures from the ADMR simulation are summarized in Table~\ref{table:YIG_SMR_SimResults}. $\rho_0$ decreases with increasing total NM/Pt layer thickness, since in these double NM structures the two parallel conducting layers both contribute to the total resistivity and since the Au and Cu layer have higher conductivity. The two-layer parallel conductance is also evidenced by the increase in $\rho_2(1\;\mathrm{T})$ compared to the YIG/Pt hybrid structure, due to the larger absolute Hall constant of Cu and Au.~\cite{henriquez_size_2010,schindler_hall_1953}
The ratio $\rho_1/\rho_0$ decreases by a factor of 3 for the YIG/Au/Pt heterostructure compared to the YIG/Pt reference sample. For the YIG/Cu/Pt heterostructure the ratio even decreases by a factor of 7. This decrease in the SMR effect can be rationalized in terms of the exponential decay of the spin current determined by the spin diffusion length in the NM and the parallel conduction channel which NM represents. A quantitative theoretical description would be highly desirable, since it would possibly allow to use SMR experiments in YIG/NM/Pt hybrids with varying NM thickness to extract the spin diffusion length in more complex structures using simple ADMR experiments in the future.

Finally, we show that the SMR effect is not limited to FMI/NM heterostructures based on the ferromagnetic insulator YIG. According to our model any FMI/NM bilayer in which the resistance of the FMI is several orders of magnitude larger than the resistance of the NM layer (such that the resistance of the NM dominates) should exhibit a SMR. To verify this conjecture we also investigated samples based on other ferromagnetic insulators and semiconductors. One sample consists of a $20\;\mathrm{nm}$ thick, (001)-oriented  magnetite (Fe$_{3}$O$_{4}$) layer on a (001)-oriented MgO substrate covered in-situ by an electron beam evaporated, $7\;\mathrm{nm}$ thin, Pt film. We note that the resistivity~\cite{reisinger_hall_2004} of Fe$_{3}$O$_{4}$ is of the order of $1\times10^{5}\;\mathrm{n\Omega m}$ at $T=300\;\mathrm{K}$, which is two orders of magnitude larger than the sheet resistivity of the Pt layer. Therefore, we performed the ADMR measurements at $T=5\;\mathrm{K}$ below the Verwey transition, where Fe$_{3}$O$_{4}$ becomes semi-insulating,~\cite{venkateshvaran_epitaxial_2009} such that the conductance is dominated by the Pt layer (Fig.~\ref{figure:YIG_SMR_Comparison}(e,f)). We note that the observed SMR ratio and the respective $\rho_i$ parameters for our Fe$_{3}$O$_{4}$/Pt heterostructure depend on the externally applied magnetic field, which is caused by the high saturation field of Fe$_{3}$O$_{4}$ due to antiphase boundaries in the film.~\cite{reisinger_hall_2004} \footnote{At $T=300\;\mathrm{K}$ we still observe the ADMR signature of the SMR superimposed on a weak AMR signal of the Fe$_{3}$O$_{4}$ layer, as the resistance of the Pt layer is still 2 orders of magnitude smaller than the resistance of the magnetite layer.} Another independent sample consists of a semiconducting, $620\;\mathrm{nm}$ thick nickel ferrite (NiFe$_{2}$O$_{4}$) layer on a (001)-oriented MgAl$_{2}$O$_{4}$ substrate with an ex-situ, sputter deposited $10\;\mathrm{nm}$ thick Pt layer (Fig.~\ref{figure:YIG_SMR_Comparison}(g,h)). The resistivity of these NiFe$_{2}$O$_{4}$ films is $1\times10^{8}\;\mathrm{n\Omega m}$ at $T=300\;\mathrm{K}$ with a bandgap of $0.5\;\mathrm{eV}$. We then performed ADMR experiments in ip, oopj, and oopt geometry as described above. In Fig.~\ref{figure:YIG_SMR_Comparison}(e)-(h) we show the corresponding results for the ip and oopj geometry. The parameters of the fits extracted from the experiments at $\mu_0H=1\;\mathrm{T}$ are summarized in Table~\ref{table:YIG_SMR_SimResults}. In all cases the fits reproduce very well the angular evolution of the data. The MR observed in experiment thus is consistent with SMR, but inconsistent with AMR. Taken together, we thus have observed the SMR effect in all FMI/Pt and FMI/NM/Pt samples studied.

\section{Conclusions}
In summary, we have quantitatively investigated the SMR effect in FMI/Pt and FMI/NM/Pt heterostructures. The SMR effect is based on the conversion of a charge to a spin current via the spin Hall effect, and back to a charge current via the inverse spin Hall effect in the NM (Pt) layer. The SMR effect characteristically depends on the absorption of the spin current at the FMI/Pt or FMI/NM interface, which in turn can be tuned via the orientation of the magnetization of the FMI with respect to the spin polarization of the spin current. Thus, the SMR effect enables a remote sensing of the magnetization direction in the FMI by simply measuring the resistance of the adjacent NM layer. We have shown that the signature of SMR is qualitatively different from conventional AMR, in particular when the magnetization has a component perpendicular to the Pt film plane. Magnetotransport measurements as a function of the magnetization orientation, rotating the magnetization from within the sample plane to a perpendicular-to-plane orientation, thus allow to disentangle AMR and SMR. We have observed SMR in YIG/Pt hybrids, as well as in YIG/Au/Pt, YIG/Cu/Pt, Fe$_3$O$_4$/Pt and NiFe$_{2}$O$_{4}$/Pt heterostructures. These results confirm the SMR thus as a universal and robust effect, that is not limited to certain material combinations. This allows to use the SMR in a wide variety of material combinations. Moreover, the excellent quantitative agreement between theory and experiment clearly show that a static magnetic proximity effect is not the origin of the observed MR in FMI/NM hybrids. We demonstrated that our quantitative model for the SMR effect allows us to extract the spin Hall angle $\alpha_\mathrm{SH}=0.11\pm0.08$ and the spin diffusion length $\lambda_\mathrm{NM}=(1.5\pm0.5)\;\mathrm{nm}$ of Pt from SMR experiments, by examining the thickness dependence of the SMR effect. This enables to use the SMR for the determination of the spin diffusion length and spin Hall angle in various NMs. The SMR is a novel, simple to measure magnetoresistance effect, which paves the way for new spin current related experiments. Last but not least, our SMR theory predicts a small AHE-like SMR contribution to the transverse resistivity due to the imaginary part of the spin mixing conductance. In our experiments we find $\rho_\mathrm{AHE}/\rho_0=1.5\times10^{-5}$ at $T=300\;\mathrm{K}$ for $t_\mathrm{Pt}=7\;\mathrm{nm}$.

\begin{acknowledgments}
Financial support via SPP 1538 "Spin-Caloric Transport" (project no. GO 944/4-1 and RE 1052/24-1), FOM (Stichting voor Fundamenteel Onderzoek der Materie), EU-ICT-7 "MACALO", the ICC-IMR, a Grant-in-Aid for JSPS Fellows, and the Nanosystems Initiative Munich (NIM) is gratefully acknowledged. The work at the University of Alabama was supported by NSF-ECCS Grant No.~1102263. M.A., S.M., M.S., S.A., M.W., H.H., S.G., M.O., R.G., and S.T.B.G.~thank A.~Erb for the preparation of the polycrystalline laser-MBE targets and T.~Brenninger for technical support.
\end{acknowledgments}
\appendix
\section{\label{Appendix}Thickness dependence of Pt resistivity}
For the simulation of the thickness dependence of the SMR ratio it is necessary to also include the thickness dependence of $\rho_\mathrm{Pt}$. The thickness dependence of the resistivity for thin films has already been extensively studied in experiment and theory.~\cite{sondheimer_mean_1952,fuchs_conductivity_1938,hoffmann_surface_1985,reiss_resistivity_1989,namba_resistivity_1970} For our analysis we use the expanded version of the Fuchs-Sondheimer theory,~\cite{sondheimer_mean_1952,fuchs_conductivity_1938} that includes a surface roughness amplitude described by the parameter $h$. In the limit that $t_\mathrm{Pt}>h$, the thickness dependence of $\rho_\mathrm{Pt}$ can be written as:~\cite{fischer_mean_1980}
\begin{equation}
\rho_\mathrm{Pt}(t_\mathrm{Pt})=\rho_{\infty} \left(1+\frac{3}{8(t_\mathrm{Pt}-h)}(l_{\infty} (1-p))\right),
\label{equ:AppendixRho}
\end{equation}
where $\rho_\infty$ is the resistivity, and $l_\infty$ the mean free path for an infinitely thick film. The parameter $p$ describes the scattering at the interfaces, for our analysis we strictly use the diffusive limit ($p=0$).
\begin{figure}[h,b,t]
  \includegraphics[width=85mm]{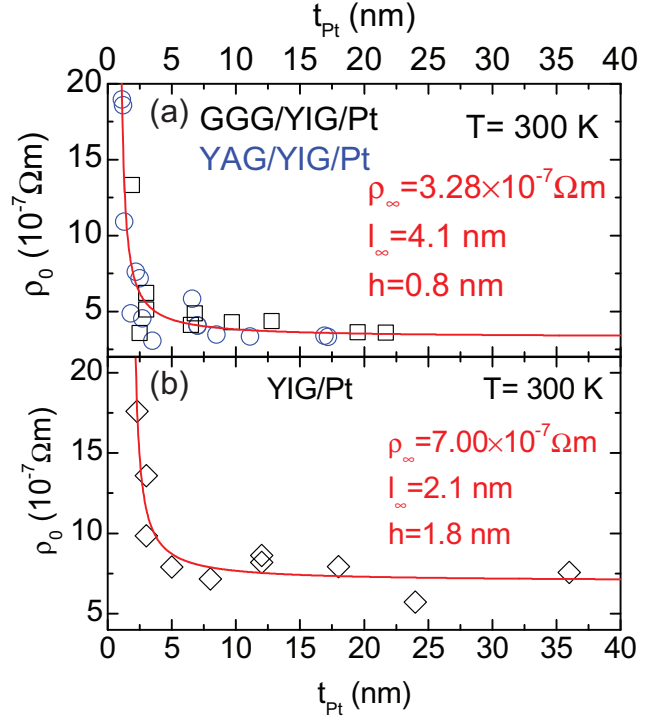}\\
  \caption[Pt thickness dependence of the resistivity]{(Color online) Evolution of the Pt resistivity as a function of Pt thickness. (a) Platinum thickness dependence of $\rho_0$ determined from different laser-MBE grown YIG/Pt ($[t_\mathrm{Pt}]\;\mathrm{nm}$) samples (open symbols) from our fits to the ADMR data (cf.~Fig.~\ref{figure:YIG_SMR_Pt_thickness}(a)). (b) Pt thickness dependence of the resistivity for liquid phase epitaxy grown YIG and sputter deposited Pt heterostrucutures (cf.~Fig.~\ref{figure:YIG_SMR_Pt_thickness}(b)). The red line in each graph is a fit to the experimental data via Eq.(\ref{equ:AppendixRho}).}
  \label{figure:rho_Pt_thickness}
\end{figure}
In Fig.\ref{figure:rho_Pt_thickness}(a) we plotted the $\rho_0(t_\mathrm{Pt})$ for different laser-MBE grown YIG/Pt ($[t_\mathrm{Pt}]\;\mathrm{nm}$) samples (cf.~Table~\ref{table:YIG_SMR_SimResults}). As expected the resistivity increases with decreasing film thickness. By fitting the experimental data by Eq.(\ref{equ:AppendixRho}) we obtain the following set of parameter: $\rho_{\infty}=(3.35\pm0.63)\times 10^{-7}\;\mathrm{\Omega m}$, $l_\infty=(4.1\pm2.0)\;\mathrm{nm}$, $h=(0.8\pm0.1)\;\mathrm{nm}$. Some experimental values deviate from the fitted theoretical curve as the assumption of a thickness independent surface roughness might be to simplistic. Nevertheless, the fitted curve reproduces the trend from our experiment and thus should be sufficient enough for the simulation of the SMR ratio. Moreover, the value for $h$ agrees nicely with the average surface roughness amplitude extracted from X-ray reflectometry measurements (cf.~Section~\ref{section_fabrication}). The results we find for the LPE grown YIG/Pt hybrids are depicted in Fig.\ref{figure:rho_Pt_thickness}(b). Here a fit to the experimental data gives: $\rho_{\infty}=(7.00\pm0.65)\times 10^{-7}\;\mathrm{\Omega m}$, $l_\infty=(2.1\pm1.1)\;\mathrm{nm}$, $h=(1.8\pm0.2)\;\mathrm{nm}$. These two sets of parameter enable us to insert Eq.(\ref{equ:AppendixRho}) into Eq.(\ref{equ:SMR_Quan_Ratio}) to describe the thickness dependence of $\rho_\mathrm{Pt}$ in our simulation in Fig.~\ref{figure:YIG_SMR_Pt_thickness}(a) and (b). Taking into account the different deposition techniques, these values compare reasonably well to published values by Fischer \textit{et al.}~\cite{fischer_mean_1980} ($\rho_{\infty}=160\;\mathrm{n\Omega m}$, $\l_{\infty}=10\;\mathrm{nm}$) for Pt thin films evaporated onto glass substrates.
\bibliography{Biblio}
\end{document}